\documentclass[preprint,unsortedaddress,superscriptaddress,onecolumn,showpacs,preprintnumbers,amsmath,amssymb,aps]{revtex4}
\usepackage{graphicx}
\usepackage{dcolumn}
\usepackage{bm}
\usepackage{latexsym}
\usepackage{color}
\usepackage{appendix}

\begin{document}

\title{Energy spectra and fluxes for Rayleigh-B\'{e}nard convection}
\author{Pankaj Kumar Mishra}
\affiliation{Department of Physics, Indian Institute of Technology, Kanpur~208 016, India}

\author{Mahendra K. Verma}
\affiliation{Department of Physics, Indian Institute of Technology, Kanpur~208 016, India}

\date{\today}

\begin{abstract}
We  compute the spectra and fluxes of the velocity and temperature fields in Rayleigh-B\'{e}nard convection in turbulent regime for a wide range of Prandtl numbers using pseudo-spectral simulations on $512^3$ grids.  Our spectral and flux results support the Kolmogorov-Obukhov (KO) scaling for zero Prandtl number and low Prandtl number ($P=0.02$) convection.  The KO scaling for the velocity field in zero-Prandtl number and low-Prandtl number convection is because of the weak  buoyancy in the inertial range (buoyancy is active only at the very low wavenumbers).  We also observe that for intermediate Prandtl numbers ($P=0.2$) the KO scaling fits better with the numerical results than the Bolgiano-Obukhov (BO) scaling.  For large Prandtl number ($P=6.8$), the spectra and flux results are somewhat inconclusive on the validity of the KO or BO scaling, yet  the BO scaling is preferred over the KO scaling for these cases.  The numerical results for $P=1$ is rather inconclusive.
\end{abstract}

\maketitle


\section{\label{sec1}Introduction}

Turbulent convection is one of the most challenging problems of classical physics~\cite{ahlers:2009}.   A large number of work on convection have been done for an idealized version called Rayleigh-B\'{e}nard convection (RBC) in which the fluid is heated between two parallel plates.  The convective flow properties depend on two nondimensional parameters:  the Rayleigh number (proportional to the buoyancy force) and the Prandtl number (the ratio of kinematic viscosity and thermal diffusivity).  The convective flow becomes turbulent when the Rayleigh number is much larger than the critical Rayleigh number. One of the important topics in the study of convective turbulence is the scaling of energy spectra and energy fluxes of the velocity and temperature fields in the inertial range.  In this paper we compute these quantities using direct numerical simulation (DNS) and compare them with the predicted values from the existing phenomenologies.

The energy spectra and fluxes for convective turbulence are more complex than those for fluid turbulence due to the presence of  the buoyancy force~\cite{siggia:1994,lohse:2010}.  For stable stratified fluid convection, Bolgiano~\cite{bolgiano:1959} and Obukhov~\cite{obukhov:1959} proposed dual cascade  in the inertial range.  For small wavenumbers (large length scale),  they predicted dominance of the buoyancy force over the inertial force leading to the velocity and temperature spectra as $k^{-11/5}$ and $k^{-7/5}$  respectively, where $k$ is the wavenumber. In this regime, the energy flux of the temperature field is constant, while the flux of the velocity field varies as $k^{-4/5}$.  For the intermediate wavenumbers, Bolgiano~\cite{bolgiano:1959} and Obukhov~\cite{obukhov:1959}  conjectured dominance of the inertial force over the buoyancy force.  Consequently the  temperature field evolves as a passive scalar, and both the velocity and temperature fields have Kolmogorov's energy spectrum ($k^{-5/3}$) and constant energy fluxes~\cite{bolgiano:1959,kolmogorov:1941,obukhov:1959}.  The length scale that separates these two different regimes of energy cascades is called the ``Bolgiano length'' ($l_{B}$). 

Later Procaccia and Zeitak~\cite{procaccia:1989}, L'vov~\cite{l'vov:1991}, and Falkovich and L'vov~\cite{falkovich:1992}  proposed the same scaling for Rayleigh-B\'{e}nard convection.  In convective turbulence, for scales above the Bolgiano length ($l>l_B$), the kinetic energy spectrum ($E^u(k)$) and the entropy spectrum ($E^{\theta}(k)$) follow the Bolgiano-Obukhov (BO) scaling  
\begin{eqnarray}
E^{u}(k)= C_{k}(\epsilon^{\theta})^{\frac{2}{5}}(\alpha g)^{\frac{4}{5}}k^{-\frac{11}{5}},\\
E^{\theta}(k)= C_{\theta}(\epsilon^{\theta})^{\frac{4}{5}}(\alpha g)^{-\frac{2}{5}}k^{-\frac{7}{5}},\\
\Pi^u(k)=C_{f}(\epsilon^{\theta})^{\frac{3}{5}}(\alpha g)^{\frac{6}{5}}k^{-\frac{4}{5}} \label{eq:Piu_4/5},
\end{eqnarray}
and for $l < l_B$, spectra follow  Kolmogorov-Obukhov (KO) scaling
\begin{eqnarray}
 E^{u}(k)= K_{ko}(\epsilon^{u})^{\frac{2}{3}}k^{-\frac{5}{3}},
 \label{KO}\\
 E^{\theta}(k)=K_{\theta}\epsilon^{\theta} (\epsilon^{u})^{-\frac{1}{3}}k^{-\frac{5}{3}}
\label{Ktheta}
\end{eqnarray}
where $\Pi^{u}$ is the kinetic energy flux, $\epsilon^{u}$ and $\epsilon^{\theta}$ are the kinetic and entropy dissipation rates respectively,  $\alpha$ is the thermal expansion coefficient of the fluid, and $g$ is the acceleration due to gravity. Note that in literature, the spectrum and the flux of the temperature field are also referred to as the ``entropy spectrum" and ``entropy flux" respectively.  

The Bolgiano length  $l_B$ has the following dependence on the convective parameters:
\begin{eqnarray}
 l_B = \frac{Nu^{\frac{1}{2}}d}{(RP)^{\frac{1}{4}}}
\label{lb}
\end{eqnarray}
where $Nu$ is the Nusselt number (dimensionless heat flux), $R$ is the Rayleigh number,  $P$ is the Prandtl number, and $d$ is the vertical height of the container.
Grossmann and L'vov~\cite{grossmann:1993} and Cioni {\em et al.}~\cite{Cioni:JFM_1997} argued that  for $P<1$, Bolgiano length is of the order of container's size.   Hence, only KO scaling is expected  in the inertial regime for low Prandtl number (low-P) convection.    For large-Prandtl number (large-P) convection, $l_{B}$ lies in the inertial regime, hence mixed scaling is expected.  Several exact relationships connecting $\epsilon^u$, $ \epsilon^\theta$, $Nu$, $R$, and $P$  have been derived for homogeneous convective turbulence.  Shraiman and Siggia~\cite{shraiman:1990} derived that
\begin{eqnarray}
 \epsilon^u & = & \frac{\nu^3}{d^4} (Nu-1) R P^{-2}   \label{eq:Piu_exact} \\
 \epsilon^\theta & = & \kappa \frac{(\Delta T)^2}{d^2} Nu \label{eq:Pitheta_exact}
\end{eqnarray}

Researchers have attempted to test the above scaling predictions [Eqs.~(1-5)] using experiments and numerical simulations (to be described later in this section).  Yet, the scaling of convective turbulence has not been conclusively established.  In a recent review, Lohse and Xia~\cite{lohse:2010} described these results critically and exhaustively. The inconsistencies of the scaling predictions with numerical and experimental results are attributed to the drastic assumptions made in the scaling arguments.  In the theory described above both thermal and viscous boundary layers are not considered appropriately.  Shraiman and Siggia~\cite{shraiman:1990} and Grossmann and Lohse~\cite{grossmann:2000} showed that the properties of the fluctuations in the boundary layer and in the bulk are rather different.  Experiments and numerical simulations reveal that the fields in the boundary layer are highly inhomogeneous and anisotropic, while the bulk flow is somewhat homogeneous and isotropic.  Hence the above scaling arguments are expected to hold only in the bulk, if at all.  The computation of the Bolgiano length $l_B$ [Eq.~(\ref{lb})] assumes uniform dissipation, which is not valid in the boundary layer. Calzavarini {\em et al.}~\cite{calzavarini:2002} have computed $l_B$ for different layers in the convective fluid; they report that $l_B/d$ is small near the walls (in the boundary layer), but $l_B/d \approx 1$ in the bulk.  In brief, the presence of boundary layers, a single $l_B$ for the whole fluid, inhomogeneity and anisotropy of the flow are some of the features that possibly make the above scaling arguments inconsistent with realistic experiments and simulations~\cite{lohse:2010}. 

To disentangle various complexities mentioned above, some researchers have idealized the geometry of RBC even further.  For example Borue and Orszag~\cite{borue:1997}, \v{S}kandera {\em et al.}~\cite{skandera:2008} considered convection in a periodic box (with thermal gradients along the vertical) and obtained KO scaling.  This feature removes the effects of the viscous and thermal boundary layers on the bulk, and hints that BO scaling is possibly due to the thermal forcing in the boundary layer~\cite{lohse:2010}.  In the present paper we consider free-slip and conducting boundary conditions in which viscous boundary layer is insignificant, while the thermal boundary layer is present.  We expect that our numerical results will suppress the effects of viscous boundary layers, and may possibly provide scaling for the bulk convective flow.

In the following discussion we briefly review the experimental studies that attempt to test the above phenomenology of RBC.  Many convection experiments measured the velocity and temperature fields only at fixed locations of the apparatus.  For such experiments ``Local Taylor hypothesis'' is invoked to relate the frequency spectrum to the wavenumber spectrum~\cite{lohse:2010,Chilla:Taylor_hypotheis}.   However, in some experiments, high resolution spatial velocity and temperature fields have been measured for computing the above mentioned spectra;  experiments by Mashiko {\em et al.}~\cite{mashiko:2004} and Sun {\em et al.}~\cite{sun:2006} belong to this category of experiments.  Chill\'{a} {\em et al.}~\cite{chilla:1993}, Zhou and Xia~\cite{zhou:2001}, and Shang and Xia~\cite{shang:2001} carried out convection experiments on water ($P \approx 7$) at large Rayleigh number and found the energy spectrum to be consistent with BO scaling.   Heslot {\em et al.}~\cite{heslot:1989} and Castaing~\cite{castaing:1990} measured frequency power spectrum of the temperature field in He gas ($0.65 < P < 1.5$) and found the spectrum to be consistent with KO scaling.  Wu {\em et al.}~\cite{wu:1990} however reported BO scaling for Helium gas through frequency spectrum measurements of temperature.   Ashkenazi and Steinberg~\cite{askenazi:1999} and Mashiko {\em et al.}~\cite{mashiko:2004} performed convection experiments for  SF$_6$ ($1 \le P \le 93$) and mercury respectively and reported the BO scaling for them.  Niemela {\em et al.}~\cite{niemela:2000} measured temperature time series in He gas and reported presence of both the KO and BO scaling. Cioni {\em et al.}~\cite{cioni:1995}  carried out experiments on mercury ($P\simeq 0.02$, a low-P fluid), and reported KO frequency spectrum for it.   Thus the outcome of these experiments are somewhat  inconclusive on the validity of the phenomenologies for RBC, yet majority appear to support the BO scaling for large-P convection, and the KO scaling for low-P scaling.

Numerical experiments provide important clues in the study of turbulence.  A series of numerical simulations of RBC have been performed to test the KO and BO scaling.  Grossmann and Lohse~\cite{grossmann:1991, grossmann:1992}  simulated RB fluid with $P=1$ under Fourier-Weierstrass approximation and reported KO scaling.  Borue and Orszag~\cite{borue:1997} and \v{S}kandera {\em et al.}~\cite{skandera:2008} performed pseudo-spectral simulation on $P=1$ fluid with periodic boundary conditions on all directions and found consistency with KO scaling.  Vincent and Yuen~\cite{Vincent:PRE_1999} performed spectral simulation for $P=1$ and $R=10^8$ using free-slip boundary conditions and reported $-5/3$ and $-3$ spectral indices for the temperature and velocity fields respectively.  They however find dual branches in the entropy spectrum.  Paul {\em et al.}~\cite{Paul:2Dspectrum} also observed dual entropy spectrum in their 2D spectral simulations with free-slip boundary conditions, albeit at lower Rayleigh numbers. Rincon~\cite{rincon:2006} performed a numerical simulation using higher order finite-difference scheme to study the effects of inhomogeneity and anisotropy on the scaling of the energy spectra; for $Ra=10^6$ and $P=1$ on $256\times\ 256 \times 128$ grids with free-slip boundary conditions, he reported that the numerical results are inconclusive in identifying a definite spectral slope.   Kerr~\cite{Kerr:1996} used pseudo-spectral method for his simulations of $P=0.7$ fluid (air) under no-slip boundary conditions and observed KO scaling.   Camussi and Verzicco~\cite{camussi:2004} performed numerical simulations for cylindrical geometry using finite difference method; they found both velocity and  temperature spectral exponents to be $-7/5$, which is inconsistent with both the KO and BO scaling.  They attribute this anomaly  to inhomogeneities and anisotropy of the flow near the boundaries.   On the whole, numerical results indicate uncertainty in the tests of the convective phenomenology.

Another way to investigate turbulent scaling is through the structure function calculations.  Following Kolmogorov, Yakhot~\cite{yakhot:1992} derived an exact analytical form for the third order structure function for  the BO scaling.   Sun {\em et al.}~\cite{sun:2006} computed the structure function of the velocity and the temperature fields using the data of their convection experiments on water and reported the KO scaling.  Kunnen {\em et al.}~\cite{kunnen:2008} performed similar calculations for Helium gas (both experiments and numerical simulation) and observed the BO scaling.     Calzavarini {\em et al.}~\cite{calzavarini:2002} computed third-order structure function using Lattice Boltzmann method for $P=1$ and reported the BO scaling.  Hence structure function studies too are inconclusive on the validity of the BO or KO scaling.

In this paper we compute  the energy spectra and cascade rates for the velocity and temperature fields  using pseudo-spectral method  on $512^3$ grids  with free-slip boundary conditions.   Our computations include zero-Prandtl number (zero-P), low-P, and large-P convection regimes  ($P=0, 0.02, 0.2, 1, 6.8$), hence we have reasonable number of numerical runs to test the convective turbulence phenomenology.   We also construct phenomenological arguments to understand zero-P and low-P numerical results.

The outline of the paper is as follows.  Section II contains the dynamical equations and the definitions of the energy spectra and fluxes.  The details and results of our numerical simulations are discussed in section  III.  We conclude in Sec. IV.


\section{Governing equations}


We numerically solve the nondimensionalized Rayleigh-B\'{e}nard equations under the Boussinesq approximation~\cite{thual:1992}
\begin{eqnarray}
\frac{\partial{\textbf{u}}}{\partial{t}}+ (\textbf{u}\cdot \nabla)\textbf{u} & = & -\nabla\sigma + R \theta \hat{z} 
+ \sqrt{\frac{P}{R}} \nabla^{2}\textbf{u},\label{eq:u}\\
P \left(\frac{\partial{\theta}}{\partial{t}}+(\textbf{u}\cdot\nabla)\theta \right) & = & u_{3} +\sqrt{\frac{P}{R}} \nabla^{2}\theta,\label{eq:T}\\
\nabla\cdot \bf{u} = 0 \label{eq:incompressible}
\end{eqnarray}
where $\textbf{u}=(u_1,u_2,u_3)$  is the velocity field, $\theta$ is the perturbations in the  temperature field from the mean temperature, $\sigma$ is the deviation of pressure from the conduction state, $R=\alpha g (\Delta T) d^3/\nu \kappa$ is the Rayleigh number, $P=\nu/\kappa$ is the Prandtl number, and $\hat{z}$ is the buoyancy direction. Here $\nu$ and $\kappa$ are the kinematic viscosity and thermal diffusivity respectively, $d$ is the vertical height of the container, and $\Delta T$ is the temperature difference between the plates.  For the nondimensionalization we have used $d$ as the length scale, $\sqrt{\alpha (\Delta T) g d}$ as the velocity scale, and $\nu (\Delta T)/\kappa$ as the temperature scale. For large-P convection, the temperature scale is taken as $\Delta T$, and the governing equations are altered accordingly.

Zero-Prandtl number (Zero-P) convection is the  limiting case of low-P convection.  The corresponding dimensionless equations for zero-P convection are
\begin{eqnarray}
 \frac{\partial{\textbf{u}}}{\partial{t}}+ (\textbf{u}\cdot\nabla)\textbf{u} & = & -\nabla\sigma +R\theta +\nabla^{2}\textbf{u}, 
 \label{eq:u_zeroP} \\
u_{3} + \nabla^{2}\theta & = & 0
\label{eq:T_zeroP}
\end{eqnarray}
 Here we use  $d$ as the length scale, $\nu/d$ as the velocity scale, and $\nu (\Delta T)/\kappa$ as the temperature scale.  
 
Boundary conditions of the systems strongly affect the properties of the convective flow~\cite{grossmann:1992,grossmann:2000}. We employ free-slip and conducting boundary conditions on the horizontal plates, hence
\begin{eqnarray}
 u_3 = \partial_{3}u_1 =  \partial_{3}u_2 = \theta = 0, ~~~~ \mbox{at}~~  z = 0, 1.  \label{bc}
\end{eqnarray}
Periodic boundary conditions are applied along the horizontal directions.  Consequently the velocity and temperature fields are expanded in terms of basis functions as
\begin{eqnarray}
u_{1,2}(x,y,z) & = & \sum_{i_x,i_y}  [ u_{1,2}(i_x,i_y, 0) + \sum_{i_z} u_{1,2}(i_x, i_y, i_z) 2 \cos(k_z z) ]
			 \exp i  (k_x x + k_y y)   \label{eq:u12basis}\\
u_3(x,y,z) & = & \sum u_3(i_x, i_y, i_z) 2 \sin(k_z z) \exp i  (k_x x + k_y y)  \\
\theta(x,y,z) & = & \sum \theta(i_x, i_y, i_z) 2 \sin(k_z z) \exp i  (k_x x + k_y y)\label{eq:thetabasis}  
 \label{eq:thetabasis}
\end{eqnarray}
where $(i_x, i_y, i_z)$ are the grid indices with $k_x = i_x \pi /\sqrt{2}$, $k_y = i_y \pi /\sqrt{2}$, and $k_z = n\pi$.

The energy spectra of the velocity field ($E^{u}(k)$) and the temperature fields ($E^\theta(k)$) are defined as 
\begin{eqnarray}
 E^u(k)= \sum_{k \le k'< k+1} \frac{1}{2} | u ({\bf k'}) |^2,  \\
 E^{\theta}(k) = \sum_{k \le k' < k+1} \frac{1}{2} |\theta({\bf k'})|^2.  
\end{eqnarray}
Here the sum is being performed over the Fourier modes in the shell $[k,k+1)$.  We will compute these spectra numerically at the steady state.  Note that the magnitude of the wavevectors in Fourier space is
\begin{equation}
k = \left( (i_x \pi /\sqrt{2})^2 +  (i_y \pi /\sqrt{2})^2 + (i_z \pi)^2 \right)^{1/2}.
\end{equation}
Here we use the fact that the aspect ratio of the box is $2 \sqrt{2}$~\cite{ahlers:2009}.

The energy flux is  a measure of the nonlinear energy transfers in turbulence~\cite{kraichnan:1959,Lesieur:book,MKV:physrep}.  The energy flux for a given wavenumber sphere  is the total energy transferred from the modes within the sphere to the modes outside the sphere.  
The energy flux for fluid and magnetohydrodynamic turbulence has been studied in great detail.  However there are only a small number of work  on the flux computations in convective turbulence~\cite{toh:1994,borue:1997,skandera:2008}.  Toh and Suzuki \cite{toh:1994} defined the kinetic energy flux $\Pi^u(k_0)$ and the entropy flux $\Pi^\theta(k_0)$  based on Kraichnan formalism~\cite{kraichnan:1959} as 
\begin{eqnarray}
\Pi^u(k_0) & = & \frac{1}{2} \sum_{k>k_0} \sum_{p,q<k_0} \delta_{\bf k,p+q} i   \frac{k_l k_m}{k_n} (1-\delta_{l,n})  \nonumber \\
		&  & \times u_l^*({\bf k})  u_m({\bf p}) u_n({\bf q}) \\
\Pi^\theta(k_0) & = & \frac{1}{2} \sum_{k>k_0} \sum_{p,q<k_0} \delta_{\bf k,p+q} i ({\bf k \cdot u(q)}) \nonumber \\ 
		   	& & \times  (\theta^*({\bf k}) \theta({\bf p})) 
\end{eqnarray}
These quantities represent the net  cascade of $|\theta|^2/2$ and $|{\bf u}|^2/2$ respectively from the modes within the wavenumber sphere of radius $k_0$ to the modes outside of the sphere.

The energy fluxes defined above  can also be defined quite conveniently using the ``mode-to-mode energy transfers" formalism discussed in Verma~\cite{MKV:physrep}.  According to  this formalism, the kinetic energy flux and the entropy flux are
\begin{eqnarray}
\Pi^u(k_0) & = &  \sum_{k \ge k_0} \sum_{p<k_0} \delta_{\bf k,p+q} \Im([{\bf k \cdot u(q)}] 
							[{\bf u^*(k) \cdot u(p)}])    \\
\Pi^\theta(k_0) & = & \sum_{k \ge k_0} \sum_{p<k_0}  \delta_{\bf k,p+q} \Im([{\bf k \cdot u(q)}] 
								[{\bf \theta^*(k) \theta(p)}])   
\end{eqnarray}
where $\Im$ represents the imaginary part of the argument.   We compute the spectra and fluxes of the velocity and temperature fields  using numerical simulations~\cite{MKV:physrep}.  These results will be described in the next section.

\section{Numerical simulations and results}

As described in the previous section, the dynamical equations of RBC are Eqs.~(\ref{eq:u}-\ref{eq:incompressible}) for low-P convection and Eqs.~(\ref{eq:incompressible}-\ref{eq:T_zeroP}) for zero-P convection.  The equations for large-P convection are similar.  We solve these equations numerically using a pseudo-spectral method under free-slip  boundary conditions for the horizontal plates, and periodic boundary conditions along the horizontal directions.   The expansion of the velocity and temperature fields are given in Eqs.~(\ref{eq:u12basis}-\ref{eq:thetabasis}).

The unidirectional initial  energy and entropy spectra for the initial conditions are of the form:
\begin{eqnarray}
 E(k)=\frac{ak^{4}}{(k^{4}+q^{4})^{1+\alpha}} \exp(-b k^{1.1}),
\end{eqnarray}
where $b=0.02$, $q=1.5$, $\alpha=2.8/12$, and $a$ as a free parameter~\cite{debliquy:2005}.  
The initial  phases are generated randomly.  Time stepping of dynamical equations are carried out using fourth-order Runge-Kutta (RK4) scheme.   We start our simulation on a smaller grid and run it until the steady state is reached. We then use the steady solution of the  lower grid as an initial condition for simulations on larger grid size at a larger $R$.  We continue this procedure until   turbulence state is reached. The final runs were performed on $512^3$ grid for $20$ large eddy turnover time on $8$ nodes and $16$ nodes of EKA, the supercomputer at Computational Research Laboratory, Pune.  Zero-P convection runs were performed on $256^3$ grid.  The $k_{max} \eta$, where $\eta$ is the Kolmogorov length, for our simulations are always greater than one indicating that our simulations are well resolved.

For the energy flux calculation, we divide the wavenumber space into 20 shells.   The first three shells are $k = (0,2)$, $[2,4)$, and $[4,8)$, and the last shell contains all modes beyond $k=568$.  Between $k=8$ and $k=568$, the wavenumber space is split into shells bounded by $[k_n, k_{n+1})$ with $k_n = 8 \times 2^{s(n-4)}$ where $s=(1/15) \ln_2{(568/8)} $.

\begin{figure}[ht]
  \begin{center}
  \includegraphics[width=1.0\columnwidth]{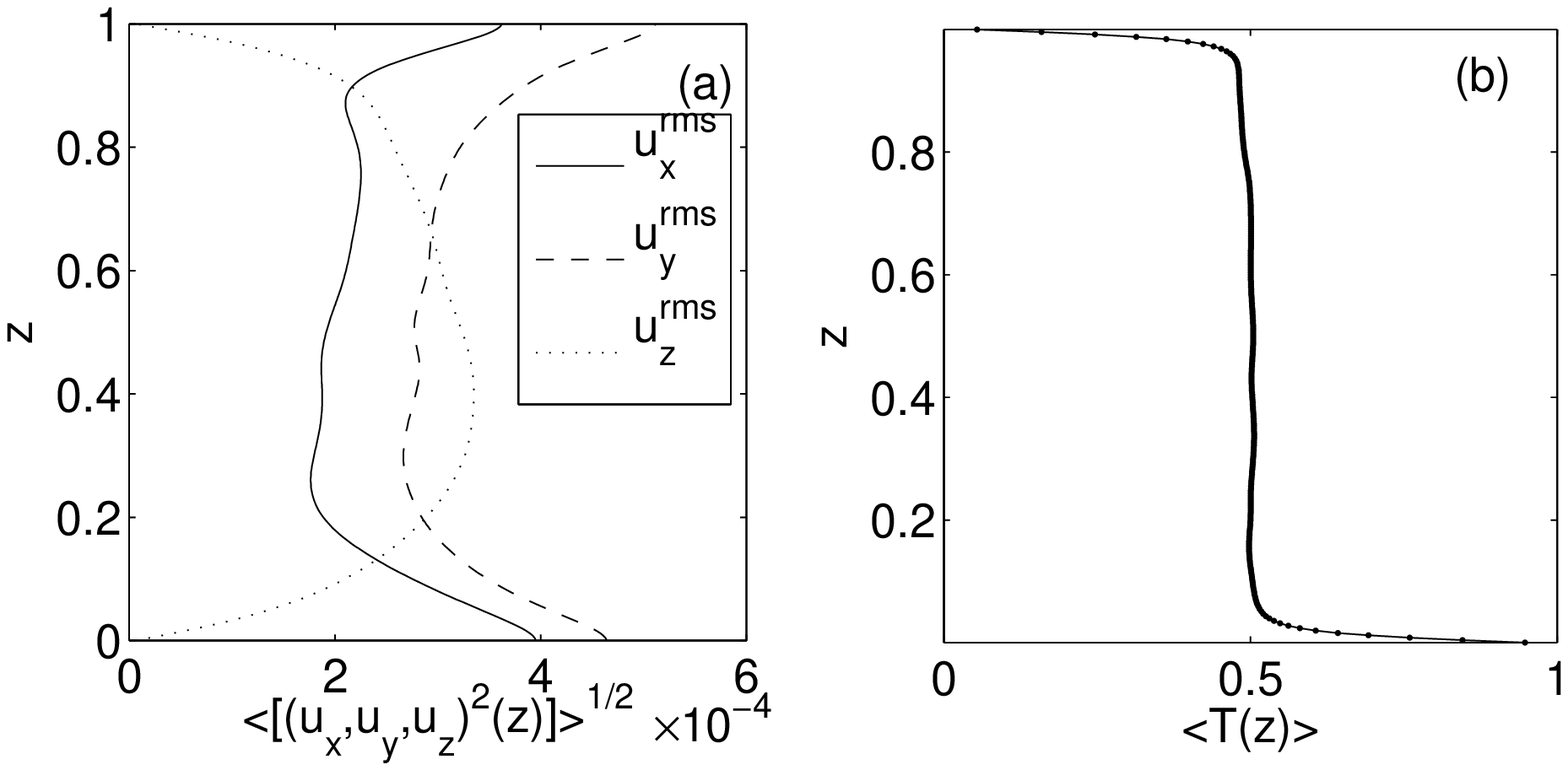}
  \end{center}
  \caption{For $R=6.6\times10^{6}$ and $P=6.8$: (a) the vertical variation of the velocity fluctuations averaged over the horizontal planes.  The solid, dashed, and dotted lines represent rms values of $u_x^{rms}$, $u_y^{rms}$, and $u_z^{rms}$ respectively; (b) the vertical variation of horizontally averaged mean temperature.}
  \label{rmsVT}
  \end{figure}

For free-slip boundary conditions, the viscous boundary layer is practically absent, while the thermal boundary layer is significant~\cite{Bala:1989}.   To probe the existence of these boundary layers we compute the average value of the rms  velocity fluctuations and the temperature field over the horizontal planes.  Figures~\ref{rmsVT}(a) and \ref{rmsVT}(b) exhibit these quantities as a function of vertical height for $R=6.6\times10^{6}$ and $P=6.8$.  We observe a thin thermal boundary layer near the horizontal plates ($\delta/d \sim 0.05 $).  The slow variation in the velocity fluctuations however demonstrates the insignificance of the viscous boundary layer.  Our results are consistent with earlier work on boundary layers~\cite{Bala:1989}.   

No-slip boundary conditions are encountered more often in convection experiments.  In our paper we are using free-slip boundary conditions for simplification.  An added advantage of the free-slip boundary conditions could be a reduction of the complexity of the viscous boundary layer; as a result, the energy spectrum of the flow may reflect the bulk properties.  Thus we may be able to probe the validity of the KO or BO scaling for the bulk flow  using these simulations.  Note that several properties of the convection are the same for both free-slip and no-slip boundary conditions, e.g., the scaling exponent of the Nusselt number vs.~Rayleigh number is the same for the two boundary conditions~\cite{Verzicco:Nusselt}.

We choose five representative Prandtl numbers $P = 0, 0.02, 0.2, 1, 6.8$ for our energy spectra and flux studies.   We compute energy spectra and fluxes, and Nusselt number using  the numerically generated data.  We also compute $\epsilon^u, \epsilon^\theta$ using the exact relationships [Eqs.~(\ref{eq:Piu_exact}-\ref{eq:Pitheta_exact})].  Kolmogorov's dissipation wavenumber ($k_d$) and ``Kolmogorov's diffusion wavenumber" ($k_c$)  are also computed using the phenomenology of passive scalar turbulence~\cite{Lesieur:book}:
\begin{eqnarray}
	k_d & = & \left(\frac{\epsilon^u}{\nu^3} \right)^{1/4}  \label{eq:kd}  \\
	k_c & = & \left(\frac{\epsilon^u}{\kappa^3} \right)^{1/4}  \label{eq:kc} \\
	\frac{k_c}{k_d} & = & P^{3/4} \label{eq:kdbykc}
\end{eqnarray}
In Table~\ref{Table1} we list the numerically computed and the estimated $\epsilon^u$ and $\epsilon^\theta$, $k_c$, $k_d$, and inverse of the Bolgiano length. The estimated values of $\epsilon^u$ and $\epsilon^\theta$ match quite well with the simulation results, thus validating our simulations. 

In the following we will discuss our numerical results on the energy spectra and fluxes for various Prandtl numbers.

\subsection{Prandtl number $P = 0$}

For $P=0$, the temperature fluctuations can be expressed as  $\theta({\bf k}) = u_3({\bf k})/k^2$ [see Eq.~$(\ref{eq:T_zeroP})$].  Consequently $E^\theta(k) \approx E^u(k)/k^4$.  Hence the entropy spectrum is very steep for zero-P  convection, and we can safely assume that the velocity field is buoyantly forced only at very large scales (small $k$).  Hence, Kolmogorov's argument for the fluid turbulence must be valid for zero-Prandtl number convection.  These arguments closely resemble the mathematical derivation of Spiegel~\cite{spiegel:1962}.  

\begin{figure}[ht]
  \begin{center}
  \includegraphics[width=1.0\columnwidth]{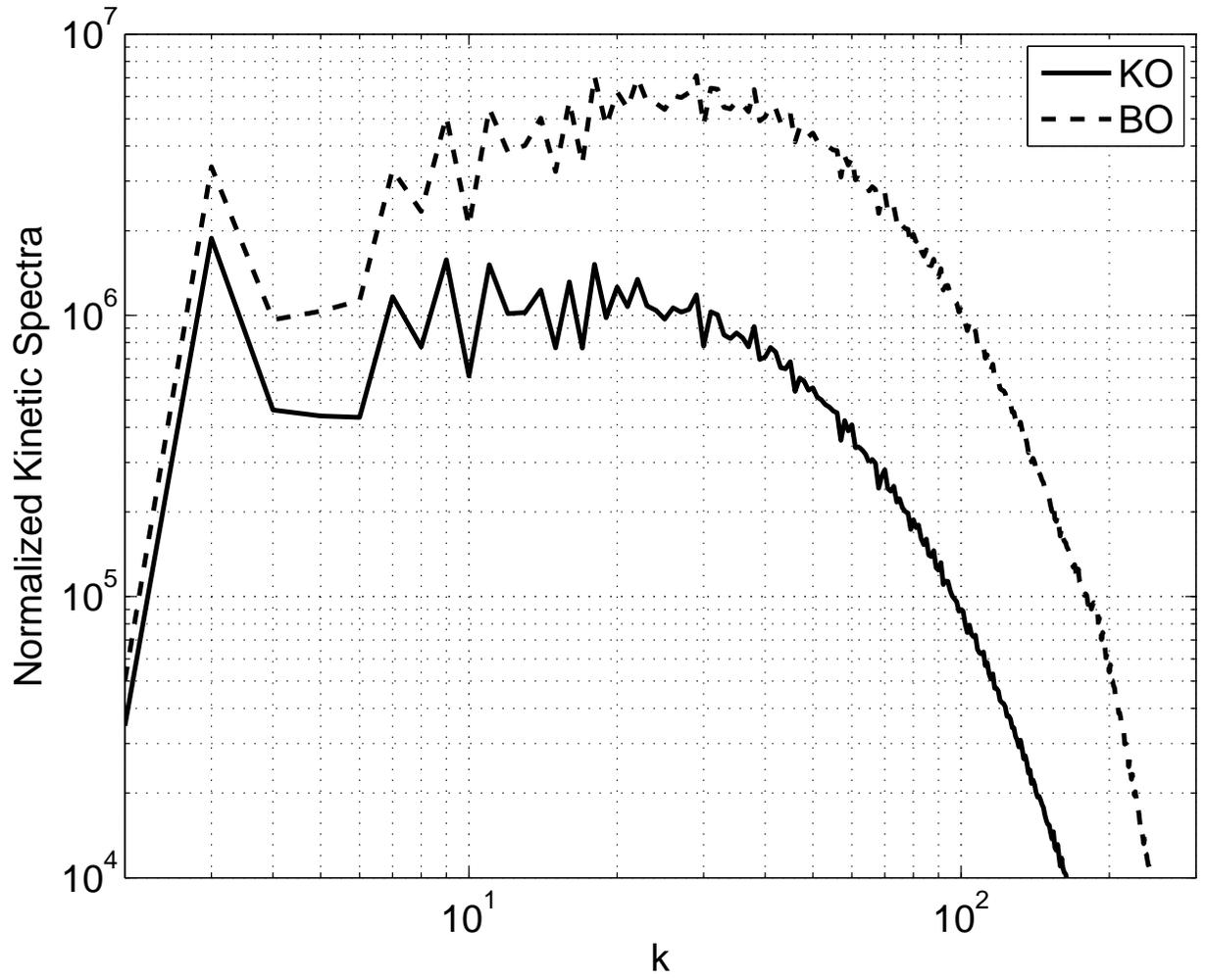}
  \end{center}
  \caption{Plot of the compensated kinetic energy spectra $E^{u}(k)k^{5/3}$ (KO) and $E^{u}(k)k^{11/5}$ (BO) vs. $k$ for $R=1.97\times10^{4}$, $P=0$ on $256^3$ grid.  The DNS spectrum matches with KO spectrum quite well.}
  \label{kespec_pr0}
  \end{figure}

\begin{figure}[ht]
  \begin{center}
  \includegraphics[width=1.0\columnwidth]{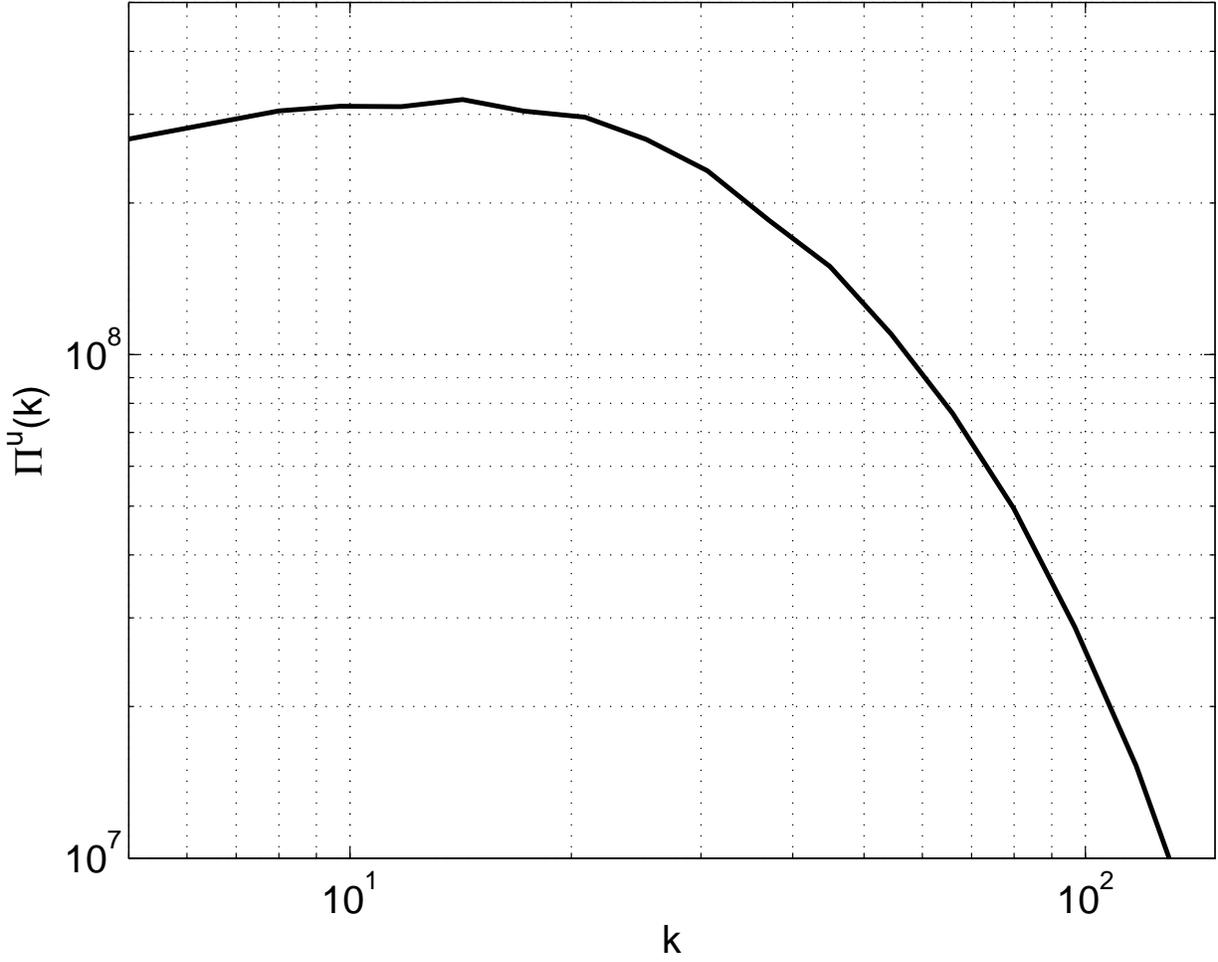}
  \end{center}
  \caption{Plot of Kinetic energy flux vs.~$k$ for $R=1.97\times10^{4}$ and $P=0$ on $256^3$ grid.  The constancy of the flux in inertial range indicates that zero-P convection follows Kolmogorov's scaling.}
  \label{flux_pr0}
  \end{figure}

We performed DNS for $P=0$ at $R=1.97\times10^{4}$ and computed the energy spectrum using the steady-state data.  In Fig.~\ref{kespec_pr0} we plot the compensated energy spectra $E(k) k^{5/3}$ (KO) and $E(k) k^{11/5}$ (BO).  Clearly the numerical plots fit better with the KO scaling than the BO scaling, thus verifying the above phenomenological arguments. Using the simulation data we also compute the kinetic energy flux that is plotted in Fig.~\ref{flux_pr0}. The kinetic energy flux is flat in the inertial range, in agreement with the KO scaling.  The Kolmogorov constant for $P=0$ is around $1.8$ (with the significant errors) which is in a reasonable agreement with the expected value of 1.6 (Kolmogorov's constant for the fluid turbulence).

In the next subsection we will discuss the numerical results for $P=0.02$ that can be a representative case for low-Prandtl number convection.

\subsection{Prandtl number $P = 0.02$}

In the previous subsection we showed  that Kolmogorov's scaling (KO) is expected to hold for zero-P convection because buoyancy for this case is dominant at very small wavenumbers.  Here we will attempt to extend the above arguments to low-Prandtl number convection.  The inertial range for the velocity and temperature fields extends almost up to the Kolmogorov dissipative wavenumber ($k_d$) and the Kolmogorov diffusive wavenumber ($k_c$) respectively.  For low-P convection, where  thermal diffusivity dominates kinematic viscosity, we expect  $k_c \ll k_d$ [see Eq.~(\ref{eq:kdbykc})].  According to the turbulence phenomenology of passive scalar turbulence, $E^{\theta}(k)$ is a power law  for $k < k_c$, and it decays exponentially for $k> k_c$.  Hence the buoyancy,  which is proportional to $\theta$ (cf. Eq.~(\ref{eq:u})), would be dominant only for low wavenumbers ($k \le k_c$), and we expect Kolmogorov's spectrum for the kinetic energy $E^u(k)$ for $ k_c < k < k_d$.  

According to Eq.~(\ref{eq:kc}), for small $\epsilon^{u}$ and large $\kappa$,  $k_c$ could be rather small.  Under such situations the above phenomenological  arguments indicate that  the velocity field follows Kolmogorov's spectrum, and the temperature field has diffusive energy spectrum.    Interestingly, the above arguments for low-Prandtl number convections are consistent with the zero-P convection for which $k_c  \rightarrow 0$ (asymptotic case).  As argued by Grossmann and L'vov~\cite{grossmann:1993}, the Bolgiano length for low-P convection could be of the order of the box size, so the BO scaling is not expected for low-P convection.

When $k_c$ is large, we need more rigorous theoretical arguments to predict the energy spectra  for $k < k_c$.  Possibly, the buoyancy term is irrelevant in ``renormalization group'' sense (see~\cite{rubinstein:1994}), and both the velocity and temperature fields may follow the KO scaling for $k<k_c$ in the inertial range.  This scenario is observed for $P=0.2$ that will be discussed in the next subsection.

In the following discussions we will compare the above phenomenological predictions with numerical results.  For $P=0.02$ we perform numerical simulation at $R=2.6\times10^{6}$ which is at the lower end of turbulent convection regime.  We compute $k_c$, $k_d$, $\epsilon^{u}$, $\epsilon^{\theta}$, $l_B^{-1}$, and energy spectra and fluxes using the numerical data.  As evident from the entries of Table~\ref{Table1}, $k_c \simeq 25$  which is much smaller than $k_d \simeq 470$.  According to the arguments given above, we expect a diffusive entropy spectrum for $k_c  < k < k_d$.  We do not expect to observe the KO scaling for $k < k_c$ since $k_c$ is too small.

\begin{figure}[h]
  \begin{center}
  \includegraphics[width=1.0\columnwidth]{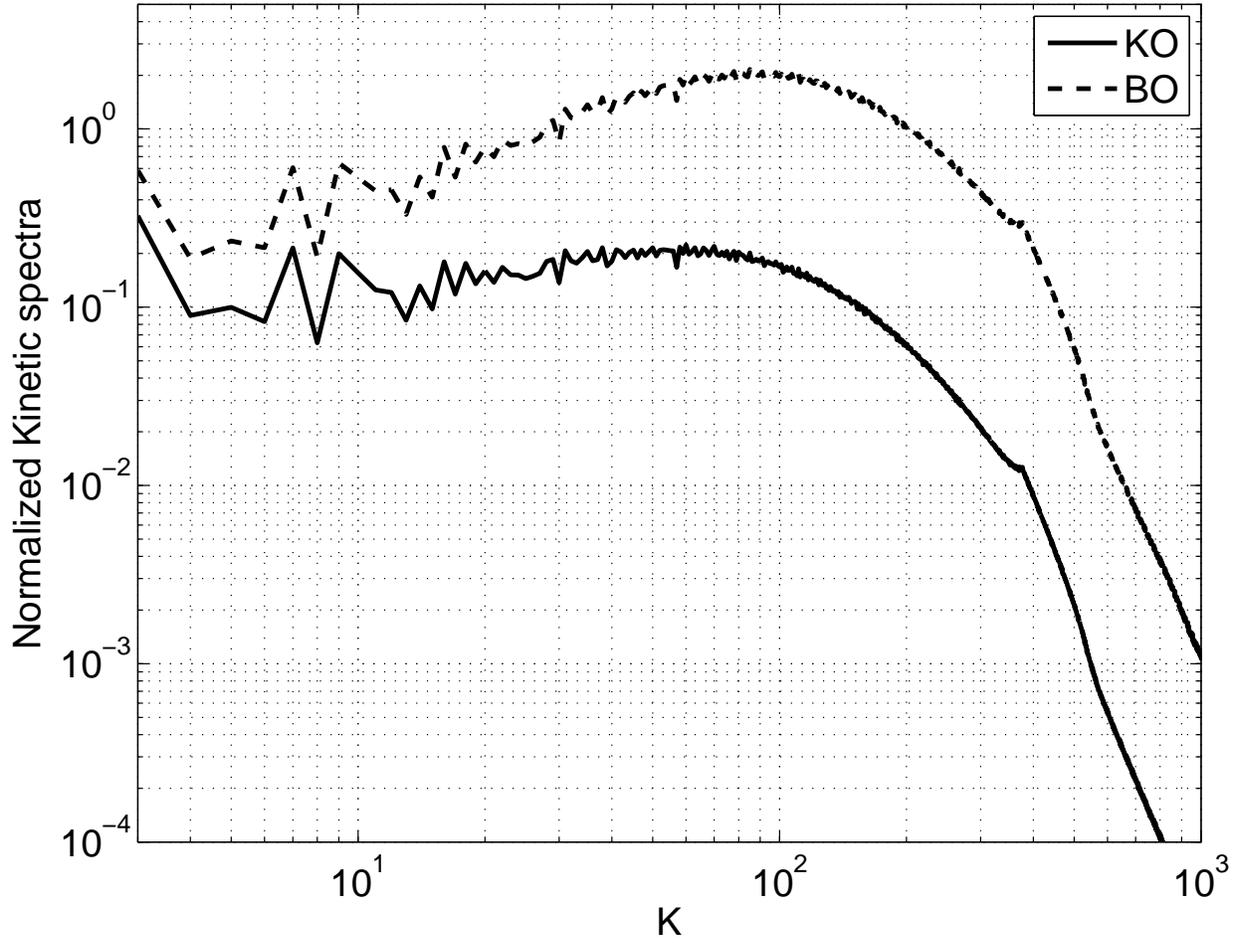}
  \end{center}
  \caption{Plot of the compensated kinetic energy spectra $E^{u}(k)k^{5/3}$ (KO) and $E^{u}(k)k^{11/5}$ (BO) vs.~$k$  for $R=2.6\times10^{6}$, $P=0.02$ on $512^3$ grid. The KO scaling is in better agreement than BO scaling.}
  \label{kespec_pr0p02}
  \end{figure}

Figure~\ref{kespec_pr0p02} contains the compensated kinetic energy spectra for the KO and BO scaling.  The KO scaling fits better with the numerical data than the BO scaling, consistent with the above phenomenological arguments.    Figure~\ref{thermspec_pr0p02}  exhibits  entropy spectrum that contains two distinct branches similar to that observed by Vincent and Yuen~\cite{Vincent:PRE_1999} and Paul {\em et al.}~\cite{Paul:2Dspectrum} in their 2D spectral simulations with similar boundary conditions as ours.  In Appendix A we construct phenomenological arguments based on energy equations and numerical results to estimate the values of the temperature modes $\theta(0,0,2n)$.  We observe that the maximum entropy transfers from the modes $\theta(n,0,n)$ and $\theta(0,n,n)$ are to the modes $\theta(0,0,2n)$ (the three indices are $i_x, i_y$, and $i_z$ respectively).  These arguments lead to predictions that $\theta(0,0,2n) \simeq -1/(2n\pi)$ and $E^\theta(2n) \simeq 1/(4n^2\pi^2)$.  For $P=0.02$ we have listed the values of $\theta(0,0,2n)$ for $n=1$ to 4 in Table~\ref{tab:theta002n}.  Here $\theta(0,0,2) \simeq -1/2\pi$, but for higher $n$'s, $|\theta(0,0,2n)| < 1/(2n\pi)$, possibly due to significant entropy transfers to other modes or due to higher thermal diffusion for low-P convection.  As we will show later, the relationshop $\theta(0,0,2n) \simeq -1/(2n\pi)$ works quite well for large-P convection.  However, a common feature borne out for all Prandtl number is that the entropy contents of $\theta(0,0,2n)$ modes are much larger that the other thermal Fourier modes, consequently yielding two branches of entropy spectrum.   

\begin{figure}[h]
  \begin{center}
  \includegraphics[width=1.0\columnwidth]{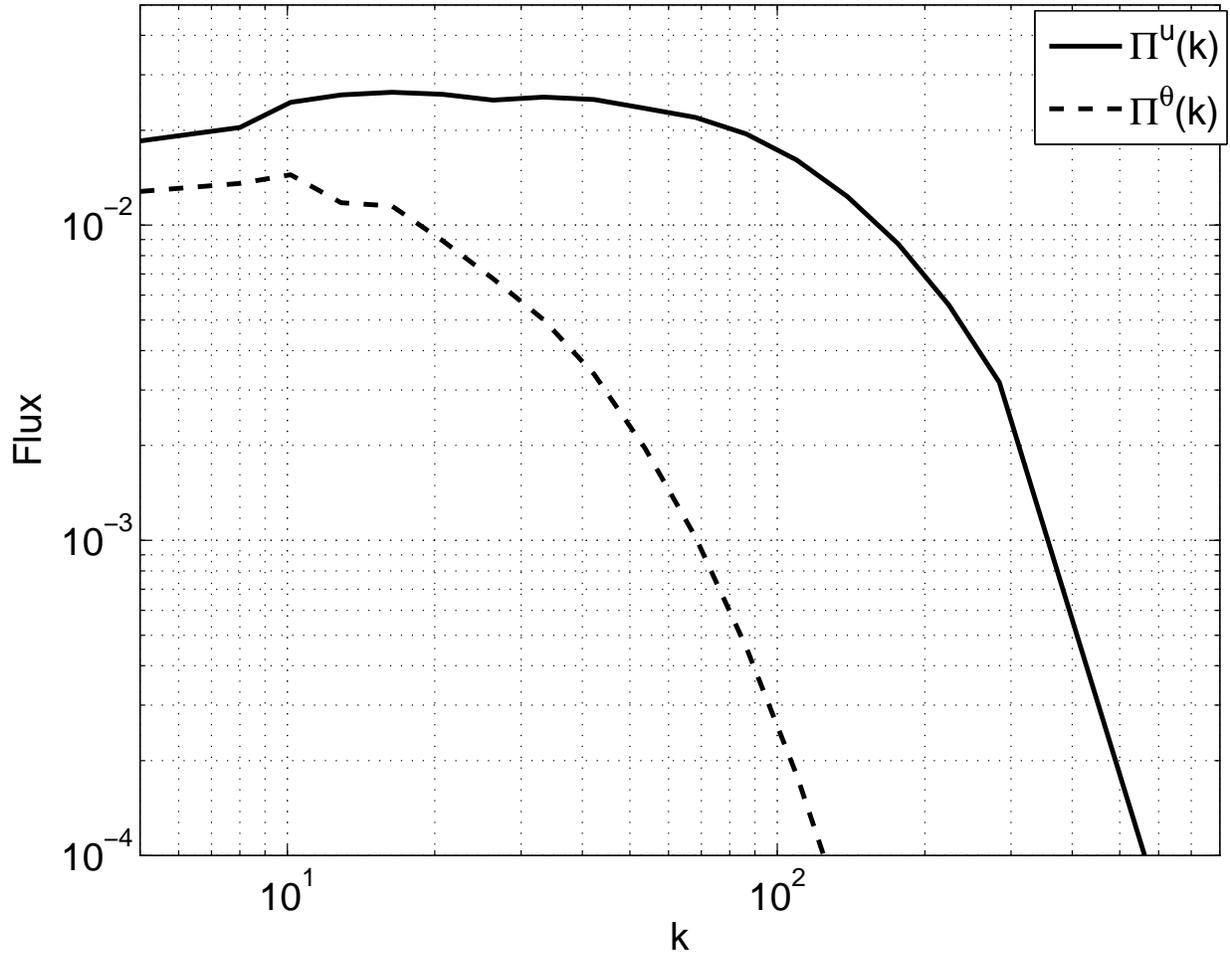}
  \end{center}
  \caption{Plot of the kinetic energy flux (solid line) and the entropy flux (dashed line) vs.~$k$ for $R=2.6\times10^{6}$, $P=0.02$ on $512^3$ grid.  The kinetic energy flux is constant in a narrow inertial range indicating agreement with the Kolmogorov's scaling for the velocity field.   The entropy flux appears to decay rather sharply suggesting diffusive entropy spectrum.  }
  \label{flux_pr0p02}
  \end{figure}

We compare the entropy spectrum with both power law and exponential fits (see Figure~\ref{thermspec_pr0p02}).  As evident from the figure, $E^\theta(k) \sim \exp(-ak)$ (the inset), which is in agreement with the phenomenological arguments given in the beginning of the subsection.    We complement our spectral analysis with energy flux studies.   Figure~\ref{flux_pr0p02} shows the kinetic energy and entropy fluxes.  The kinetic energy flux is  flat for more than a decade indicating Kolmogorov's spectrum for the velocity field, in agreement with the KO scaling for the velocity field.  The entropy flux however drops sharply, consistent with the exponential nature of the entropy spectrum.   Using Eqs.~(\ref{eq:Piu_exact},\ref{eq:Pitheta_exact}) we compute $\epsilon^u$ and $\epsilon^\theta$ that are quite close to the numerically computed energy and entropy fluxes (see Table~\ref{Table1}).  Also, the numerical estimate of $k_c$ and $k_d$ are in general agreement with the spectra and flux plots. 

On the whole, the numerical results for $P=0.02$, which is a representative of low-P convection, appear to favour KO scaling  for the velocity field.  The temperature spectrum appears to be diffusive for the most wavenumber region.   These numerical results are in good agreement with the phenomenological arguments presented above for low-P convection.

In the next subsection we report energy spectra and fluxes for $P=0.2$.

\subsection{Prandtl number $P = 0.2$}

\begin{figure}[ht]
  \begin{center}
  \includegraphics[width=1.0\columnwidth]{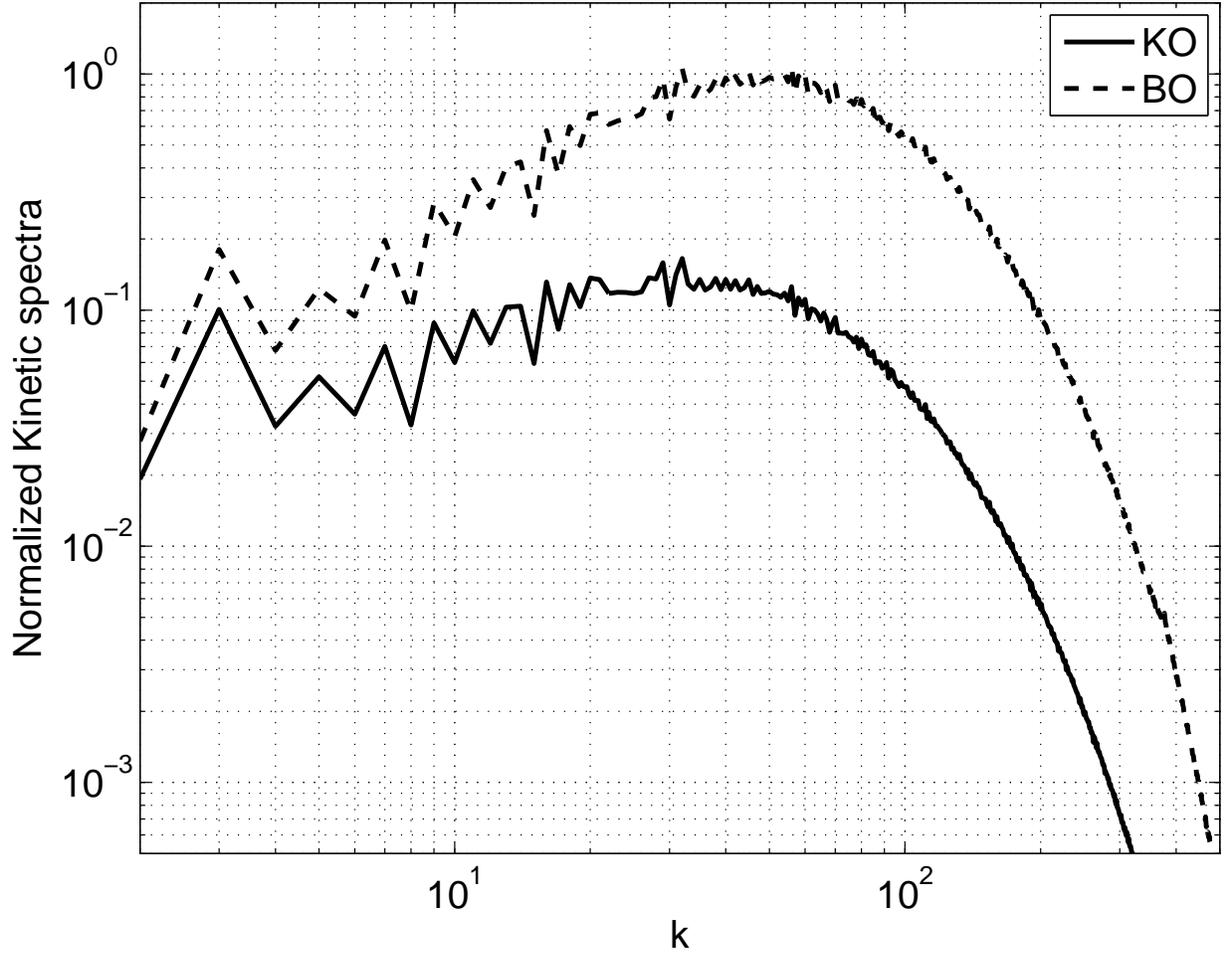}
  \end{center}
  \caption{Plot of the compensated kinetic energy spectra $E^{u}(k)k^{5/3}$ (KO) and $E^{u}(k)k^{11/5}$ (BO) vs.~$k$ for $R=6.6\times10^{6}$, $P=0.2$ on $512^3$ grid. The numerical results match better with the KO scaling than the BO scaling.}
  \label{kespec_pr0p2}
  \end{figure}

Next we present our numerical results for $P=0.2$ at $R=6.6\times10^{6}$.  In Fig.~\ref{kespec_pr0p2} we plot  the compensated kinetic energy spectra $E^u(k) k^{5/3}$ (KO) and $E^u(k) k^{11/5}$ (BO).  Even though both the BO and the KO scaling do not fit very well with the numerically computed energy spectrum, yet the KO scaling is in better agreement with the numerical data than the BO scaling.   

In Fig.~\ref{thermspec_pr0p2} we plot the entropy spectrum, which has significant inertial range. Note that $k_c =68$ (see Table ~\ref{Table1}).  We obtain bi-spectra similar to that for $P=0.02$.  As described in the Appendix A, the upper curves represent the spectrum of the Fourier modes $\theta(0,0,2n)$, and it matches reasonably well with $k^{-2}$ spectrum.  As evident from the entries of Table~\ref{tab:theta002n}, $\theta(0,0,2n)$ matches with $-1/(2n\pi)$ within a factor of 2.  The lower curve however appears to fit better with the KO scaling than the BO scaling.   Note that the upper curve of the entropy spectrum contains small number of Fourier modes, hence the nonlinear energy transfers from these modes may be insignificant.  There are large number of Fourier modes associated with the lower branch of  $E^\theta(k)$, and the energy flux possibly results from the nonlinear interactions among these modes.   For this reason we compare the lower branch of the entropy spectrum to either KO or BO scaling.

\begin{figure}[h]
  \begin{center}
  \includegraphics[width=1.0\columnwidth]{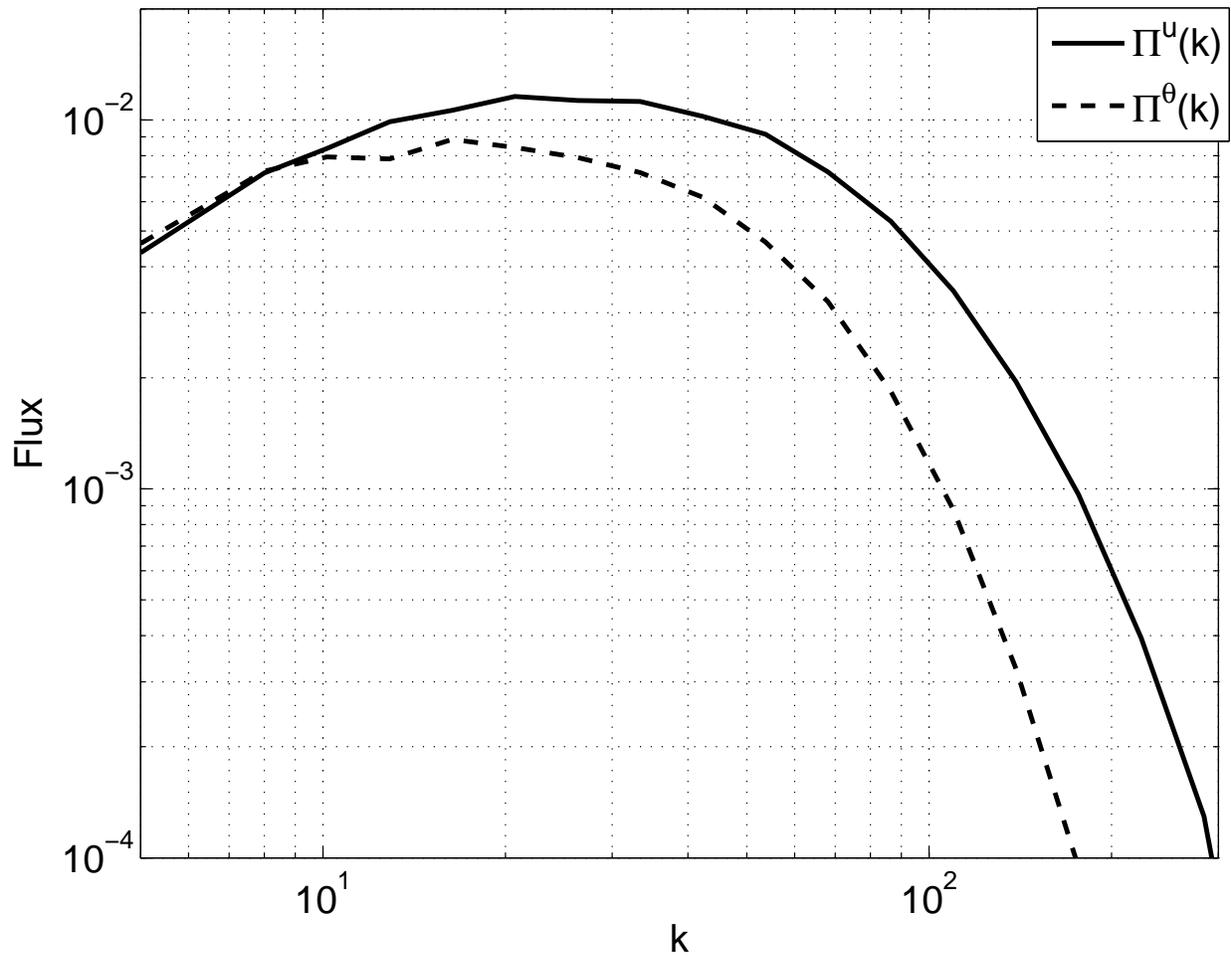}
  \end{center}
  \caption{Plot of the kinetic energy flux (solid line) and the entropy flux (dashed line) vs.~$k$ for $R=6.6\times10^{6}$, $P=0.2$ on $512^3$ grid.  The kinetic energy and entropy fluxes are constant in the narrow inertial range indicating a general agreement with the KO scaling.}
  \label{flux_pr0p2}
  \end{figure}

Next, we compute the energy fluxes for the velocity and temperature fields for the same run. We observe constant fluxes for both the velocity and temperature fields as exhibited in Fig.~\ref{flux_pr0p2}.  Thus both energy spectra and flux results appear to favor the KO scaling more than the BO scaling.  Given the kinetic energy spectrum and flux (in the common inertial range), we compute Kolmogorov's constant using Eq.~(\ref{KO}) that yields $K_{Ko} \approx 2.0$ with significant error.  Considering the uncertainties in the numerical fits, this value is  in a reasonable agreement with  Kolmogorov's constant for the fluid or the passive-scalar turbulence measured earlier using experiments and numerical simulations.

We also compute $\epsilon^u$, $\epsilon^\theta$, $k_d$, and $k_c$ using Eqs.~(\ref{eq:Piu_exact},\ref{eq:Pitheta_exact},\ref{eq:kd},\ref{eq:kc}).  These numbers are listed in Table ~\ref{Table1}.  The predicted values of $\epsilon^u$ and $\epsilon^\theta$ are in general agreement with the simulation results.  We observe that $k_c < k_d$, which is also evident in the spectra and flux plots.   An important point to note is that $k_c \sim 68$ is rather large.  Hence the arguments presented in the earlier subsection for low-P convection will not hold here.  More rigorous arguments are required to understand the phenomenology for $P=0.2$.

After considering $P=0.2$, we turn to convection for $P=1$ .

\subsection{Prandtl number $P = 1$}

\begin{figure}[ht]
  \begin{center}
  \includegraphics[width=1.0\columnwidth]{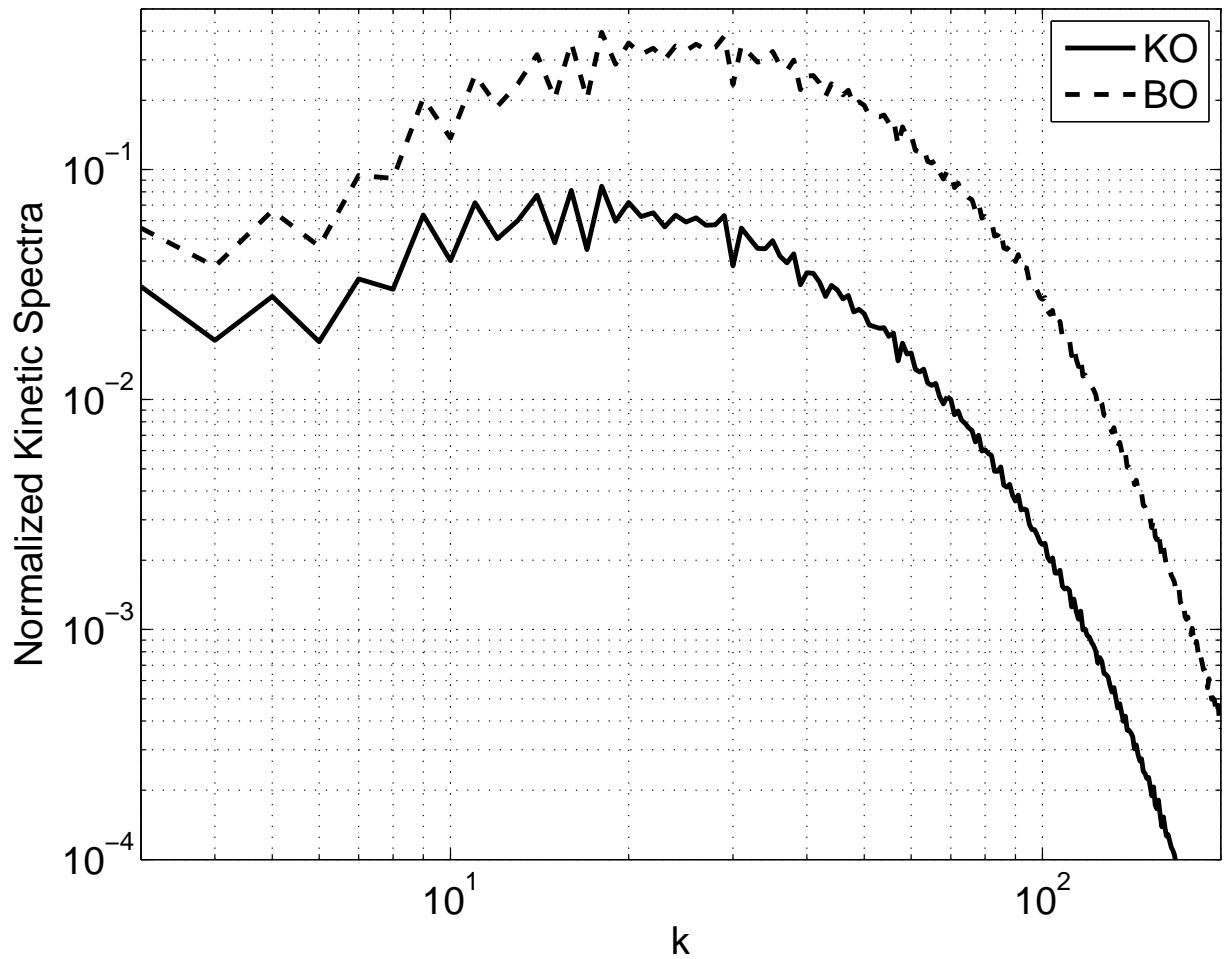}
  \end{center}
  \caption{Plot of the compensated kinetic energy spectra $E^{u}(k)k^{5/3}$ (KO) and $E^{u}(k)k^{11/5}$ (BO) vs.~$k$ for $R=6.6\times10^{6}$, $P=1$ on $512^3$ grid.  We cannot infer which phenomenology fits better with the plots. }
  \label{kespec_pr1}
  \end{figure}

Next we present the energy spectra and fluxes for the kinetic energy and entropy for $P=1$ at $R= 6.6 \times 10^6$.  Figures~\ref{kespec_pr1} and  \ref{thermspec_pr1} exhibit the compensated kinetic energy spectra and entropy spectrum respectively.  The kinetic energy spectra plots are inconclusive since both the compensated plots for the KO and BO scaling are equally flat, albeit at different wavenumber ranges.  

\begin{figure}[ht]
  \begin{center}
  \includegraphics[width=1.0\columnwidth]{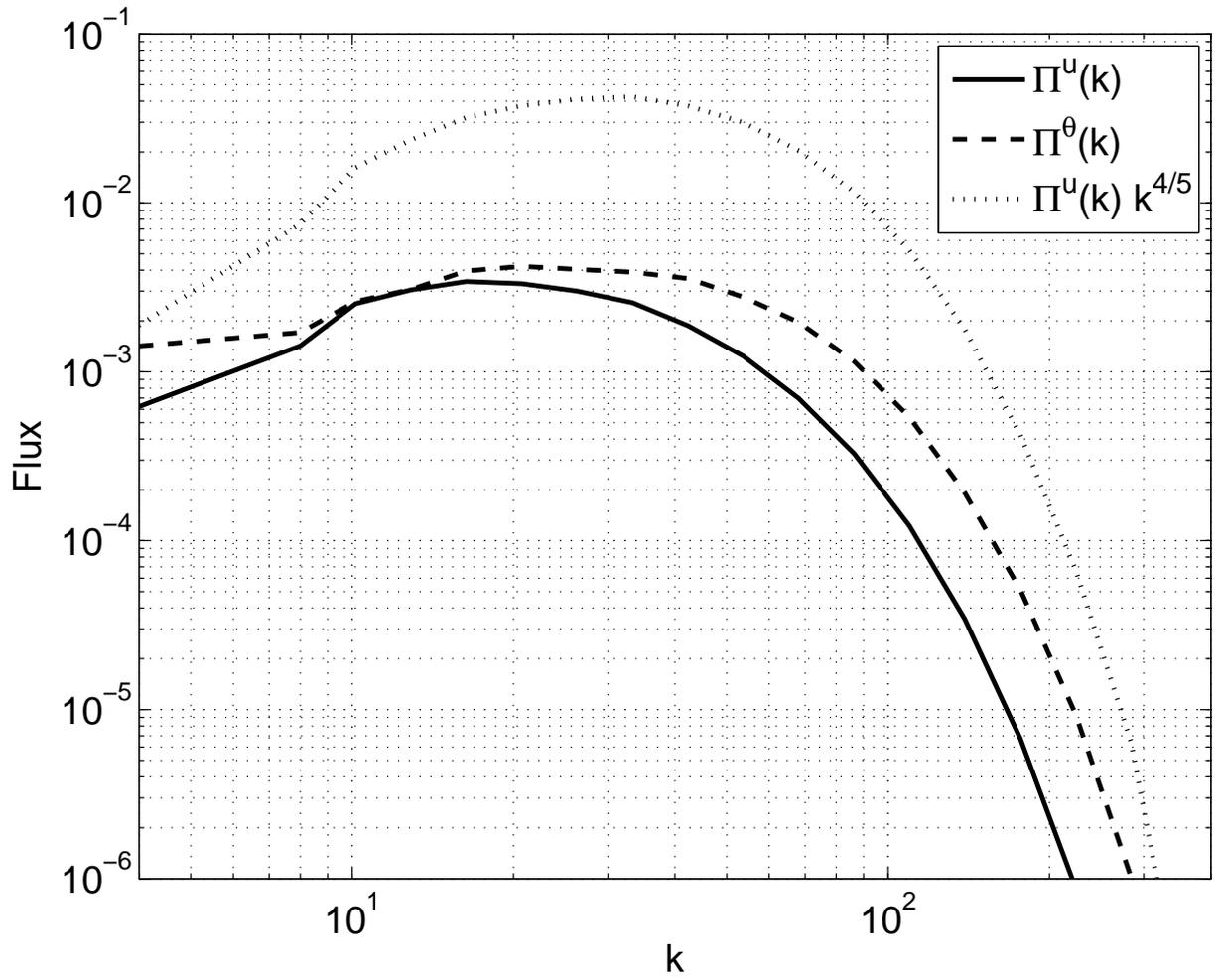}
  \end{center}
  \caption{Plot of the kinetic energy flux (solid line) and the entropy flux (dashed line) vs.~$k$ for $R=6.6\times10^{6}$, $P=1$ on $512^3$ grid. The dotted line represents $\Pi^u(k) k^{4/5}$ curve.  The flux results are inconclusive about the nature of scaling.}
  \label{flux_pr1}
  \end{figure}

The entropy spectrum, shown in Fig.~\ref{thermspec_pr1}, has two distinct branches similar to low-P cases.   In agreement with the arguments of Appendix A, the upper  branch of the entropy spectrum follows $E^\theta(2n) \sim n^{-2}$.   A comparison of  the lower curve with the BO or KO scaling indicates that neither of the scaling fits well with the numerical data.  Figure~\ref{flux_pr1} shows the kinetic energy and entropy fluxes along with the compensated kinetic energy flux $\Pi^{u} k^{4/5}$.  The flux plots are also inconclusive.

Overall, the numerical results for $P=1$ are rather inconclusive.  The inverse Bolgiano length for $P=1$ is approximately 9.0 (see Table 1), hence the phenomenologies predict BO scaling for $k< 9\pi \sim 28$ and KO scaling for $28 < k < k_d$.  Clearly the wavenumber range of BO or KO scaling is too small to be able to infer any scaling.  Also, the arguments put forth for the validity of the KO scaling for low-P convection based on the dominance of buoyancy force for low wavenumbers cannot be extended to $P \ge 1$.

In the next subsection we discuss the simulation results on convective turbulence for $P=6.8$.

\subsection{Prandtl number $P = 6.8$}

\begin{figure}[ht]
  \begin{center}
  \includegraphics[width=1.0\columnwidth]{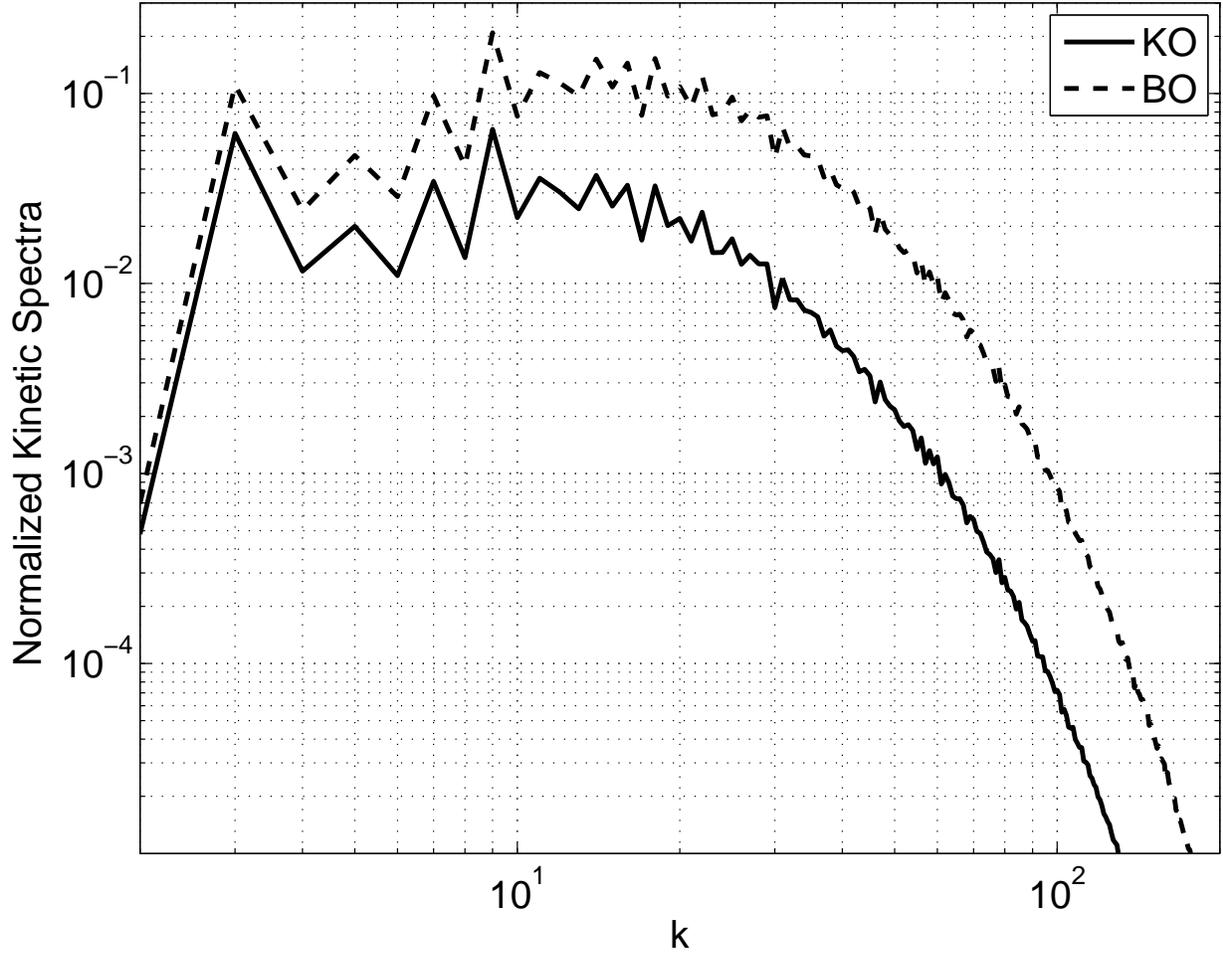}
  \end{center}
  \caption{Plot of the compensated kinetic energy spectra $E^{u}(k)k^{5/3}$ (KO) and $E^{u}(k)k^{11/5}$  (BO) vs. $k$ for $P=6.8$, $R=6.6\times10^{6}$ on $512^3$ grid.   The fit is somewhat inconclusive, yet the BO scaling appears to fit better with the numerical data than the KO scaling.}
  \label {kespec_pr6p8}
  \end{figure}

At last, we present the kinetic energy spectrum for $P=6.8$ at $R=6.6 \times 10^{6}$.   In Fig.~\ref{kespec_pr6p8} we plot  the compensated kinetic energy spectra  $E^u(k) k^{5/3}$ (KO) and $E^u(k) k^{11/5}$ (BO).  The flat regions in both the plots are rather short, yet the BO line appears to be in a better agreement with the numerical results than the KO line.  The inverse of Bolgiano length $l_B^{-1}$  is around $15.0$. Hence according to the convective turbulence phenomenology discussed in Section I, the BO scaling should hold for $k< \pi l_B^{-1}$, and the KO scaling should hold for $k> \pi l_B^{-1}$.  The BO scaling appears to be present in our numerical results, but the KO scaling is not observable.  The dominance of dissipation for modes with $k> \pi l_B^{-1}$  in our $512^3$ simulation may be a reason for the absence of the KO scaling.  We need higher resolution simulation to investigate this issue.

Figure~\ref{thermspec_pr6p8} exhibits  dual branches in the entropy spectrum similar to those discussed earlier. The upper spectral curves representing the  modes  $\theta(0,0,2n)$ follow $k^{-2}$ scaling as predicted in Appendix A.  The Fourier modes $\theta(0,0,2n) \simeq -1/(2n\pi)$ as evident from Table~\ref{tab:theta002n}.  For the lower branch, both the KO and BO scaling are not in good agreement with the entropy spectrum, yet the BO scaling fits better with the numerical data than the KO scaling. 

\begin{figure}[ht]
  \begin{center}
  \includegraphics[width=1.0\columnwidth]{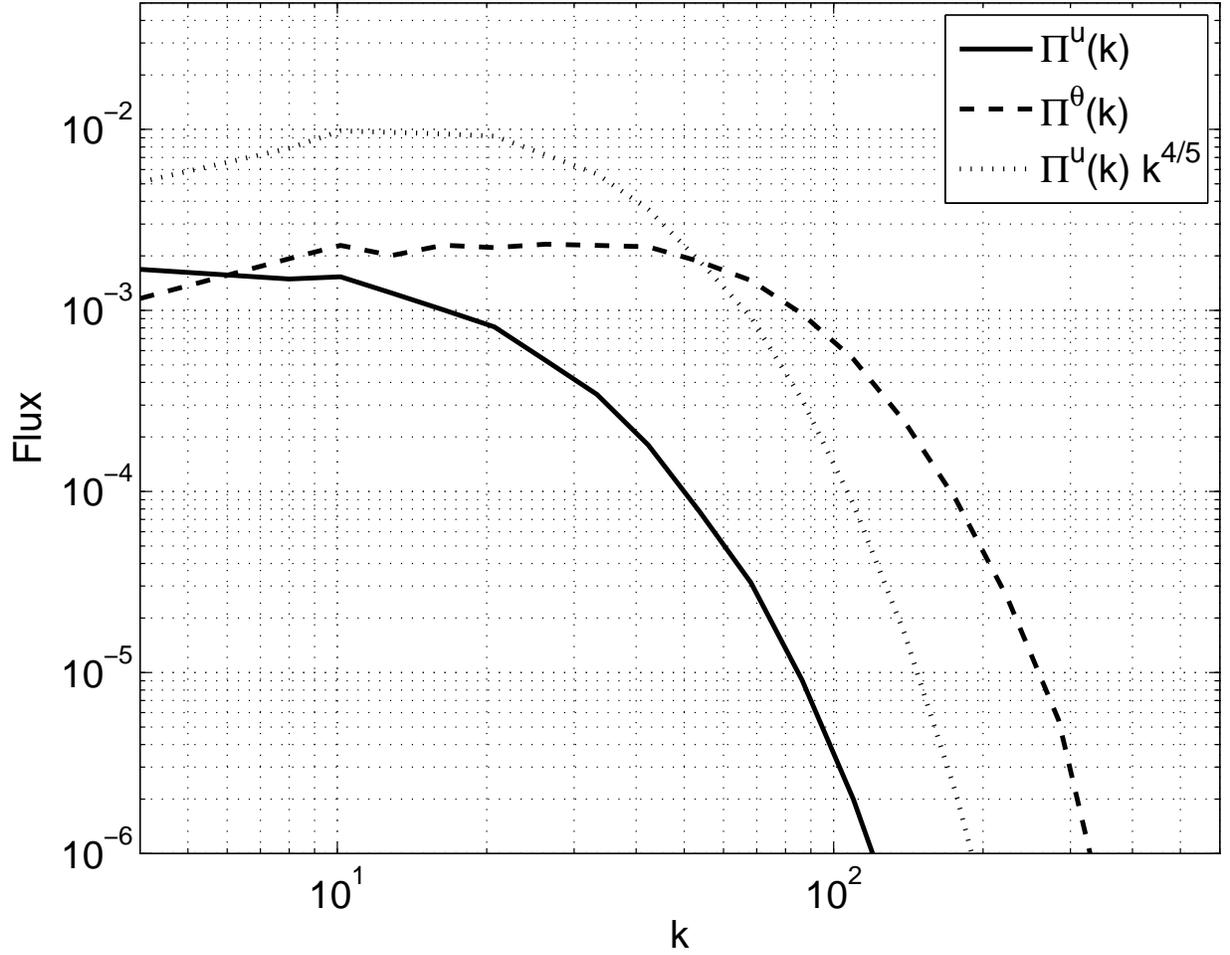}
  \end{center}
  \caption{Plot of the kinetic energy flux (solid line) and the entropy flux (dashed line) vs. $k$ for $R=6.6\times10^{6}$, $P=6.8$ on $512^3$ grid. The normalized kinetic energy flux (multiplied by $k^{4/5}$) is also shown in the figure as a dotted line.  $\Pi^{u} k^{4/5}$ and $\Pi^{\theta}$ are constant for a range of wavenumbers.  } 
  \label{flux_pr6p8}
  \end{figure}

 Recall that for large-P convection, under the BO scaling, the entropy flux is constant but the energy flux varies as $k^{-4/5}$ (see Eq.~(\ref{eq:Piu_4/5})). In contrast, in the KO scaling the fluxes of the kinetic energy and the entropy are constant.  In Fig.~\ref{flux_pr6p8} we plot both the fluxes as well as the compensated kinetic energy flux $\Pi^{u}(k) k^{4/5}$.  We observe that $\Pi^u(k)$ falls rather steeply as a function of wavenumbers, but the compensated kinetic energy flux is constant in a narrow band of the inertial range.   The  entropy flux is also a constant for a significantly large wavenumber range.  Thus the flux results tend to favor BO scaling for $P=6.8$.  
  
Our numerical results on the energy spectra and fluxes are somewhat inconclusive, but the BO scaling scaling appears to fit better with the simulation results.

\section{Conclusions}

We numerically compute the spectra and fluxes of the velocity and temperature fields of convective turbulence using a pseudo-spectral method.  We performed these simulations for a large range of Prandtl numbers---zero-P, low-P, and large-P.   The Rayleigh number of our simulation is around a million, which is at the lower end of turbulent convection.   We apply free-slip and thermal boundary conditions for our simulations.  As a result, the viscous boundary layer is rather weak, but the thermal boundary layer is quite significant.  Consequently,   our numerical results possibly reflect the scaling for the bulk convective flow.  The simulation results of kinetic energy and entropy fluxes are in good agreement with their estimates using exact relations, thus validating our numerical simulations.

We find that for nonzero Prandtl numbers, the entropy spectrum shows dual branches.  Our simulation results indicate that the maximum entropy transfer from the modes $\theta(n,0,n)$ and $\theta(0,n,n)$ are to the mode $\theta(0,0,2n)$. These observations combined with entropy evolution equations yield $\theta(0,0,2n) \simeq -1/(2n\pi)$ and $E^\theta(2n) \simeq 1/(4n^2\pi^2)$.  For large-P convection, these predictions fit very well with the upper branch of the entropy spectrum.  The upper branch however has only a small number of modes, and they probably do not contribute significantly to the entropy flux.  For this reason we compare the lower branch of the entropy spectrum to either the Kolmogorov-Obukhov (KO) or the Bolgiano-Obukhov (BO) scaling.

For zero-P convection, the temperature field is active only for very small wavenumbers since $E^{\theta}(k) \sim E^u(k)/k^4$. Hence, buoyancy is active only for very small wavenumbers leading to Kolmogorov's scaling just like in fluid turbulence ($E^u(k) \sim k^{-5/3}$ and $\Pi^{u}(k) \sim const$).  We observe such behaviour in our numerical simulation. 

The above phenomenological arguments for zero-P convection can be extended to low-P convection.  For this case, Kolmogorov's diffusive wavenumber $k_c$ is much smaller than Kolmogorov's dissipation wavenumber $k_d$.  Hence, the temperature field will be diffusive for $k>k_c$, and the forcing due to buoyancy is active only for low wavenumbers ($k<k_c$).  Consequently, we expect Kolmogorov's spectrum for the velocity field for $k_c < k <k_d$.   We numerically compute the energy spectra and fluxes for $P=0.02$, and observe diffusive spectrum for the temperature field and Kolmogorov's spectrum for the velocity field.  Thus the phenomenological arguments presented above are in agreement with our simulations.  For $k<k_c$, the inertial range is too narrow to ascertain any of the KO or the BO scaling.  A large Rayleigh number simulation could possibly resolve the scaling in this range.

We have also computed the spectra and fluxes for $P=0.2$.  For this case, the Kolmogorov-Obukhov (KO) scaling appears to fit better than the Bolgiano-Obukhov (BO) scaling with the energy spectra and fluxes of the velocity and temperature fields ($E^u(k) \sim k^{-5/3}$, $E^{\theta} \sim k^{-5/3}$, $\Pi^{u}(k) \sim const$, and $\Pi^{\theta}(k) \sim const$).   Numerical results for $P=1$ are inconclusive regarding the phenomenology.  Simulations results for $P=6.8$, which is a sample of large-P convection, too are inconclusive, however the BO scaling appears to fit better than the KO scaling in this case.   $P=1$ and large-P convection require more refined simulations for resolving these issues.

When we compare our results with earlier experiments and simulations, we observe general agreement with the findings of Cioni {\em et al.}~\cite{cioni:1995} where they reported KO scaling for mercury ($P=0.02$, low-P). Chill\'{a} {\em et al.}~\cite{chilla:1993},  Zhou and Xia~\cite{zhou:2001}, and Shang and Xia~\cite{shang:2001} performed experiments on water and reported the BO scaling for it. Our simulation results are in general agreement with the the above experimental results.   A word of caution is in order:  our simulations use free-slip boundary conditions that differs from the no-slip boundary conditions of the experiments.  Also, realistic convective flows are quite complex due to the presence of boundary layers, anisotropic forcing (buoyancy), plumes, large-scale circulation (LSC) etc. all of which have not been analyzed carefully in our analysis.  Several past numerical simulations and experiments have attempted to study these features~\cite{camussi:2004}.  Our emphasis in this paper has been on the bulk energy spectrum and fluxes.  Note that the plumes and LSC typically affect the low-wavenumber regime of the energy spectrum, and may not significantly affect the inertial-range isotropic energy spectra being investigated in the present paper.

In summary, we observe the KO scaling for zero-P and low-P convection in our numerical simulations.  For large-P convection, the numerical results are not very convincing, yet the BO scaling matches with the numerical results better than KO scaling.  These results are in general agreement with some of the earlier experimental and numerical results.
We provide phenomenological arguments to support KO scaling for low-P and zero-P convection.  More rigorous theories like renormalization group analysis and very high resolution simulations could be very useful in providing further insights into this complex problem.    Unfortunately convective turbulence simulations beyond $512^3$ are prohibitively expensive at this stage.   Also, more complex features like inhomogeneity, anisotropy, boundary layers need to investigated.   Future experiments, simulations, and theoretical modeling will hopefully resolve this outstanding problem.

\appendix
\section{Entropy Spectrum}

The entropy spectrum exhibits dual branches.  In this appendix we discuss the reasons for this behaviour.  We start with the entropy equation for the $\theta(n,0,n)$ mode, which is
\begin{equation}
\frac{\partial}{\partial{t}}\frac{|\theta(n,0,n)|^2}{2} = T^{\theta}(n,0,n)+\Re[u_{3}(n,0,n)\theta^{*}(n,0,n)]-\frac{1}{\sqrt{PR}}(n^2 \pi^2+ n^2 k_c^2)|\theta(n,0,n)|^2
\label{eq:theta_n0n}
\end{equation}
where $T^\theta(n,0,n)$ is the nonlinear entropy transfer to the mode $\theta(n,0,n)$, and $k_c = \pi/\sqrt{2}$.  The second term in the RHS is the entropy production rate $P^\theta(n,0,n)$ due to the vertical velocity, and the last term provides the dissipation  rate of entropy due to thermal diffusivity.  The entropy equation for the $\theta(0,0,2n)$ mode is very similar.   We compute $T^{\theta}(n,0,n)$ and $P^\theta(n,0,n)$ from the simulation data, and find these quantities to be highly variable.  Yet we compute them at a given instant of time in the steady state regime.  In this regime, $\partial |\theta(n,0,n)|^2/\partial t \simeq 0$, and the dissipation term is also quite small.   In Table~\ref{tab:Ttheta} we list the numerical values of $T^{\theta}(n,0,n)$, $T^{\theta}(0,n,n)$, $P^\theta(n,0,n)$, and  $P^\theta(0,n,n)$ at an instant.  Clearly, $T^\theta \simeq -P^\theta$ indicating that the entropy generated by $u_3$ is transferred to the higher modes by nonlinear transfer.

From Eq.~\ref{eq:theta_n0n} we can conclude that the $\theta(n,0,n)$ mode gains energy through the entropy production term~($P^\theta(n,0,n)$), and loses energy to other modes through nonlinear entropy transfer~($T^\theta(n,0,n)$).  When we compute the energy transfers functions explicitly, we find that the dominant entropy transfer from $\theta(n,0,n)$ is to the $\theta(0,0,2n)$.  The ``mode to mode energy transfer'' formalism~\cite{MKV:physrep} provides us the entropy transfer rate from $\theta(0,0,2n)$ to the $\theta(n,0,n)$ with  $\bf{u}$$(-n,0,n)$ acting as a mediator, which is 
\begin{eqnarray}
S({\bf k|p|q}) & = & -\Im\{ 2 n \pi (-i) u_3(-n,0,n) \theta(n,0,n) \theta(0,0,2n) \} \nonumber \\
&= &  2 n \pi \theta(0,0,2n) \Re[u_3^*(n,0,n) \theta(n,0,n)] 
\end{eqnarray}
with ${\bf k}=(n,0,n)$, ${\bf p}=(0,0,2n)$, ${\bf q}=(-n,0,n)$. The term $\Im()$ stands for the imaginary part of the arguments. 
The above formula has been adopted from the mode-to-mode energy transfer formulas for Fourier basis to mixed basis used for the free-slip boundary conditions [Eqs.~(\ref{eq:u12basis}-\ref{eq:thetabasis})].

\begin{figure}[h]
  \begin{center}
  \includegraphics[width=1.0\columnwidth]{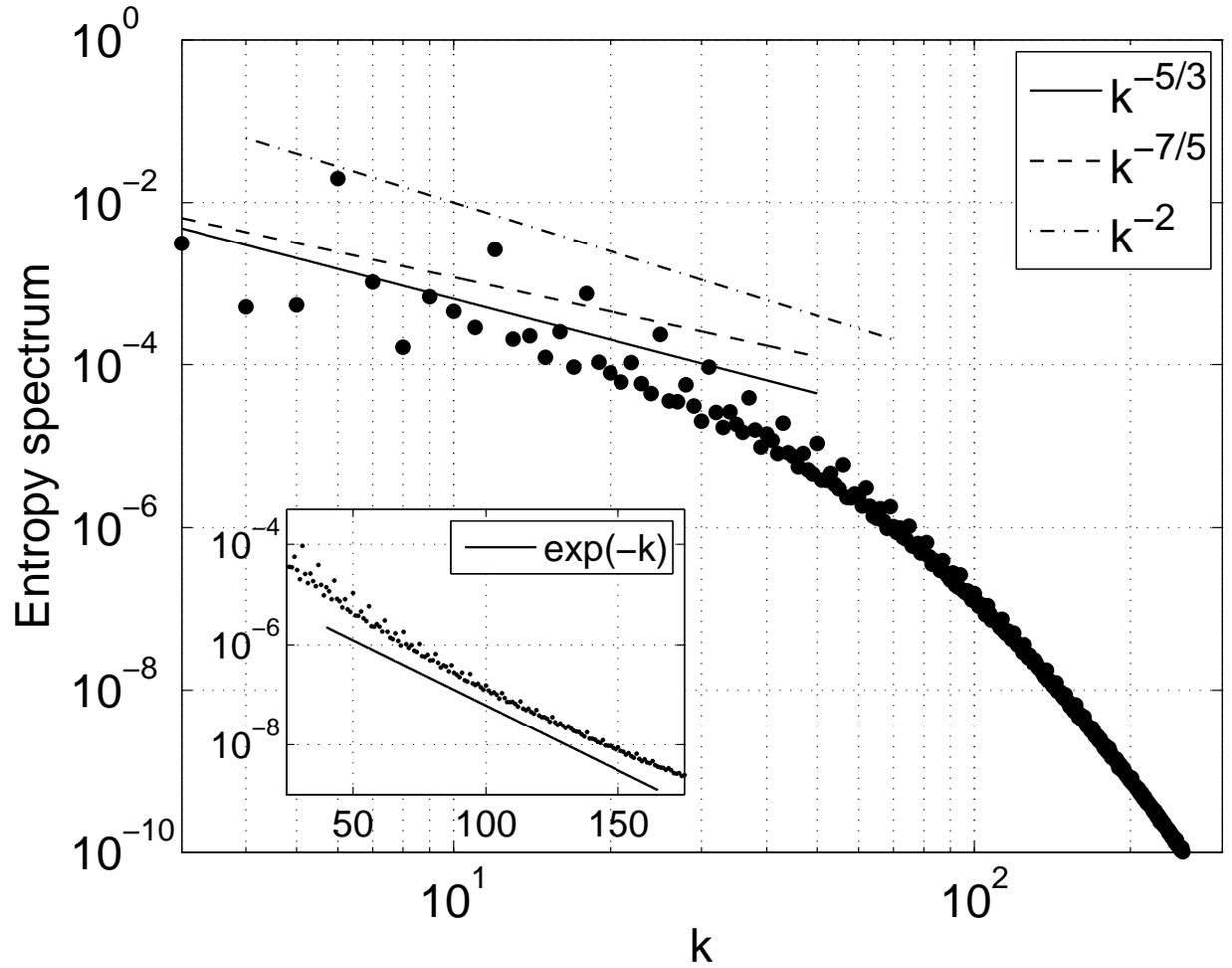}
  \end{center}
  \caption{Plot of the entropy spectrum  for $R=2.6\times10^{6}$, $P=0.02$ on $512^3$ grid.  The exponential fit in the inset indicates the diffusive nature of the entropy spectrum.  The upper part of the entropy spectrum corresponds to the $\theta(0,0,2n)$ modes. }
  \label{thermspec_pr0p02}
  \end{figure}

\begin{figure}[h]
  \begin{center}
  \includegraphics[width=1.0\columnwidth]{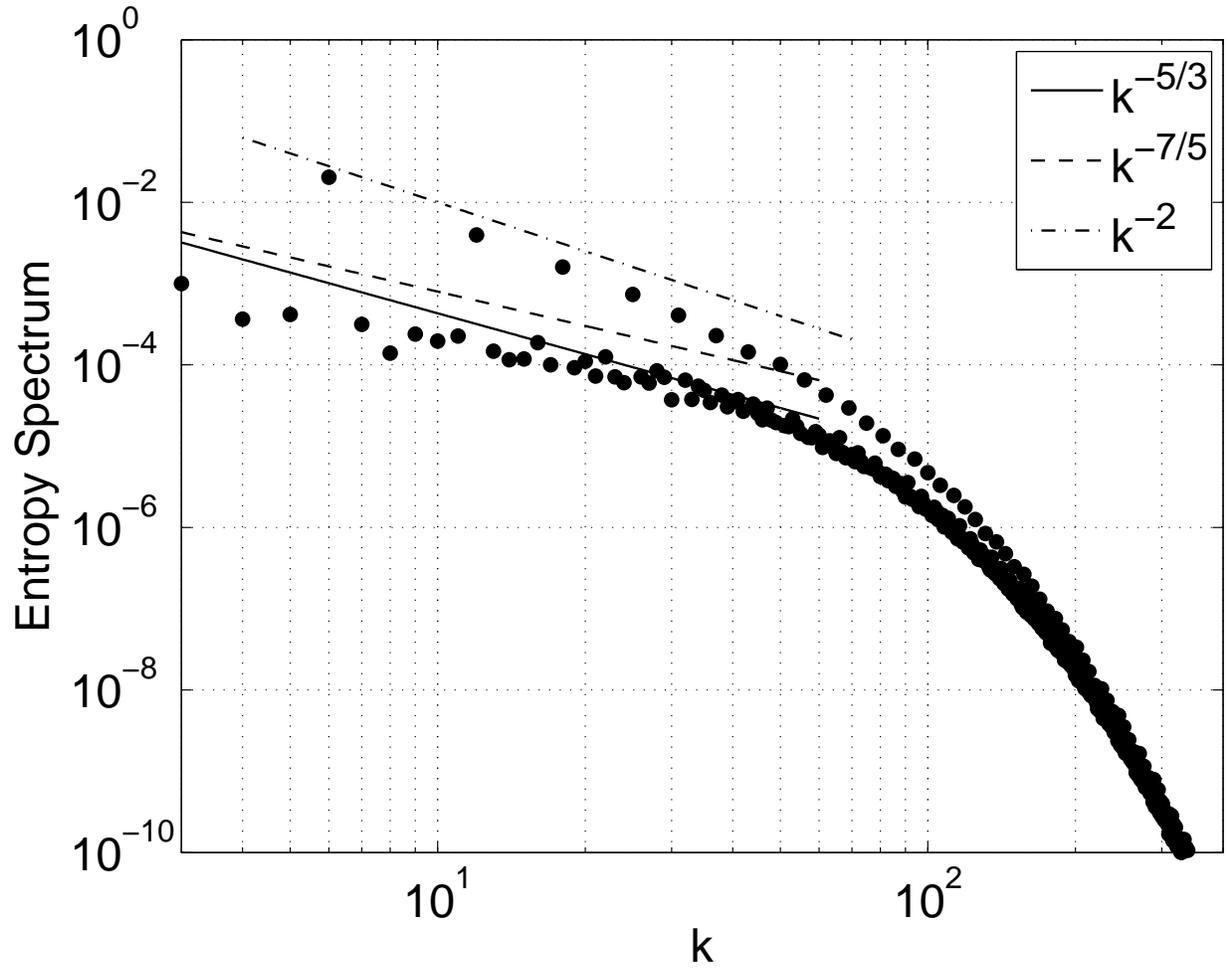}
  \end{center}
  \caption{Plot of the entropy spectrum vs.~$k$ for $R=6.6\times10^{6}$, $P=0.2$ on $512^3$ grid.  
The lower part of the entropy spectrum matches better with the KO scaling than the BO scaling.  The upper part of the entropy spectrum corresponds to the $\theta(0,0,2n)$ modes, and it is in general agreement with $k^{-2}$ fit.  }
  \label{thermspec_pr0p2}
  \end{figure}

\begin{figure}[ht]
  \begin{center}
  \includegraphics[width=1.0\columnwidth]{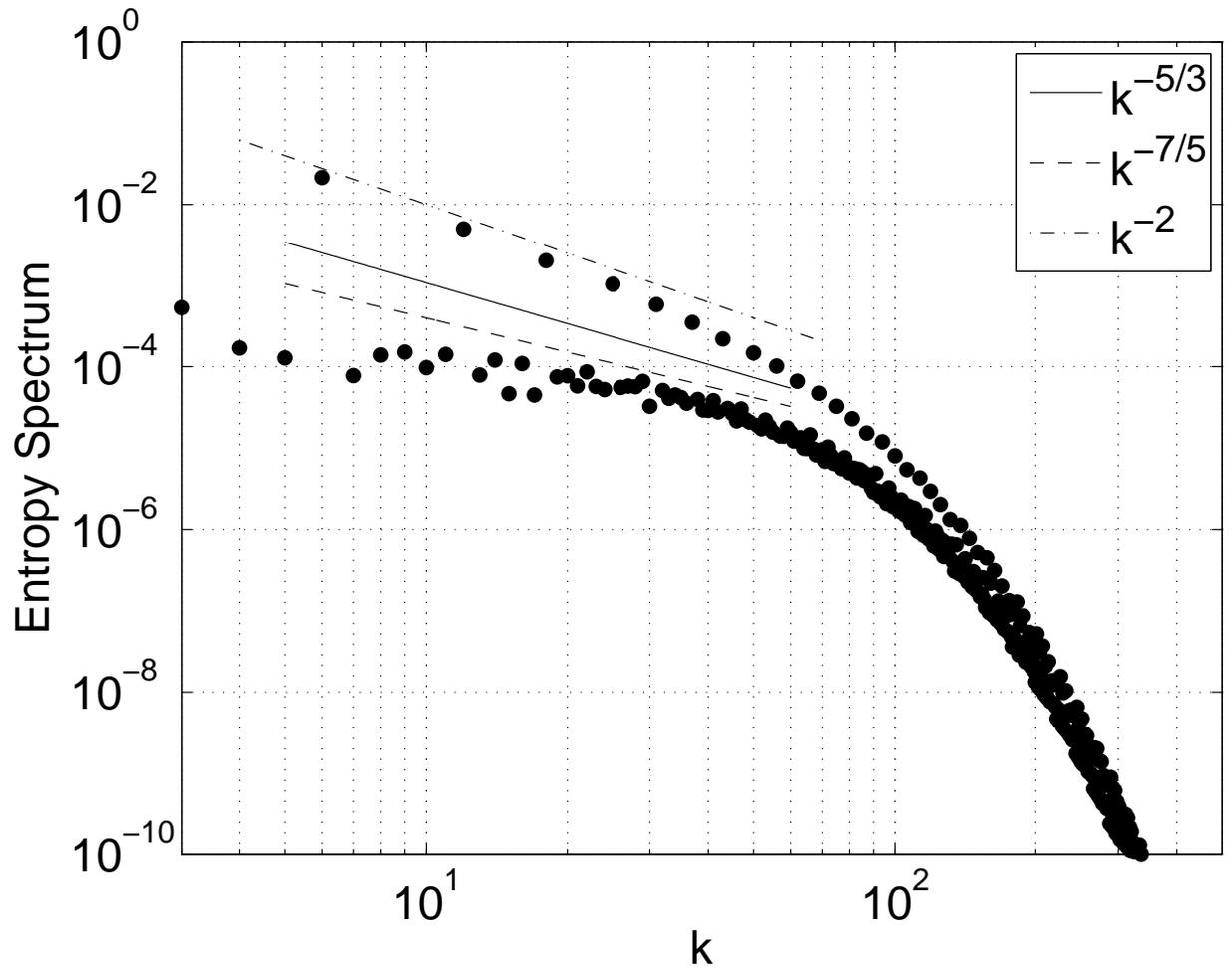}
  \end{center}
  \caption{Plot of the entropy spectrum vs.~$k$ for $R=6.6\times10^{6}$, $P=1$ on $512^3$ grid.  Both the KO and BO lines do not fit with the lower branch of the entropy spectrum. The upper part of the entropy spectrum matches with  $k^{-2}$ quite well. }
  \label{thermspec_pr1}
  \end{figure}

\begin{figure}[ht]
  \begin{center}
  \includegraphics[width=1.0\columnwidth]{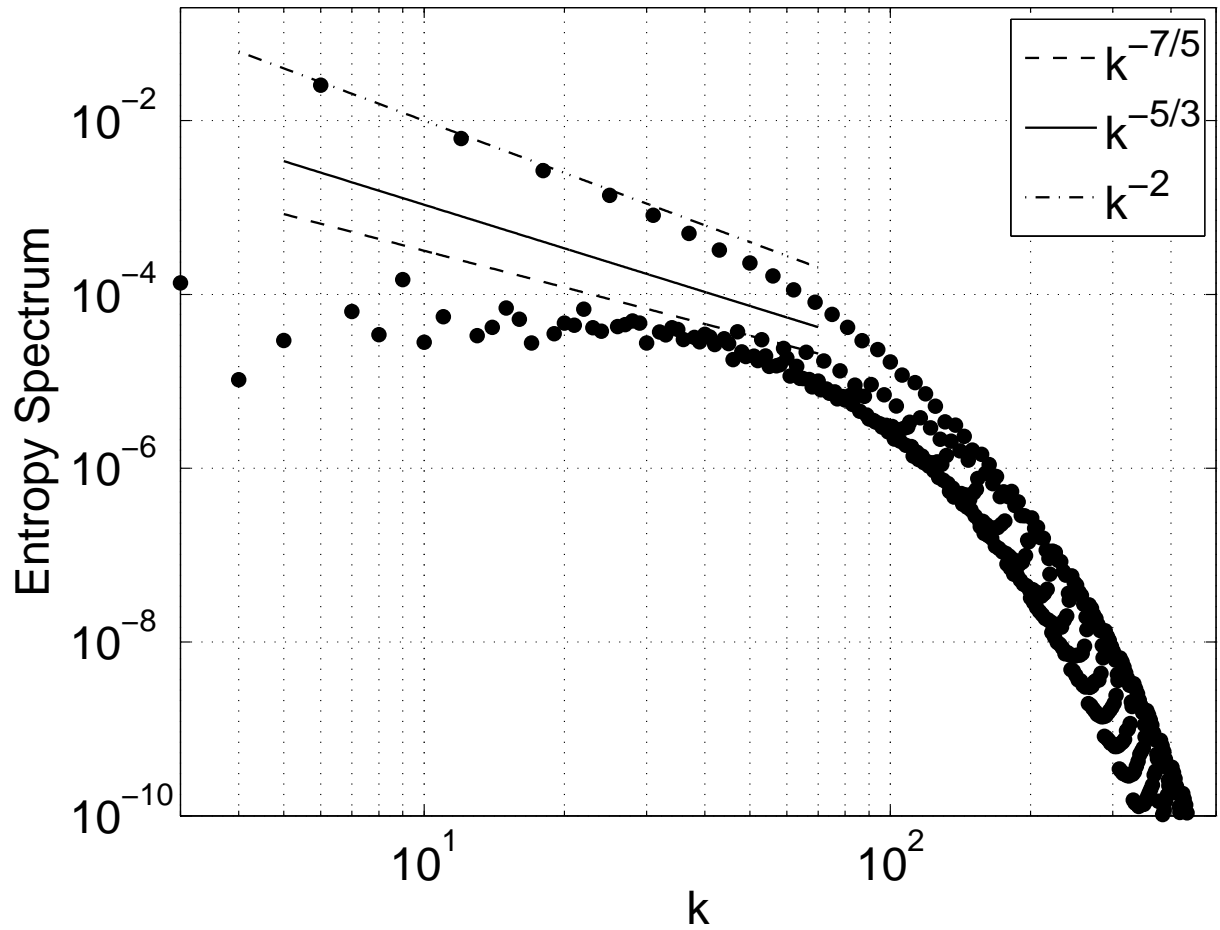}
  \end{center}
  \caption{Plot of entropy spectrum $E^{\theta}(k)$ vs. $k$ for $P=6.8$, $R=6.6\times10^{6}$ on $512^3$ grid.   Even though both the KO and BO lines do not fit well with the data, BO line is closer to the lower part of the spectrum.  The upper branch matches with $k^{-2}$ quite well.  }
  \label{thermspec_pr6p8}
  \end{figure}

We compute $S({\bf k|p|q})$ using the numerical data at the same instant of time when we compute $T^\theta(n,0,n)$, and compare it with $T^\theta(n,0,n)$ and the entropy production.  As evident from the entries of the Table~\ref{tab:Ttheta}
\begin{equation}
T^\theta(n,0,n) \simeq S(n,0,n|0,0,2n|-n,0,n) \simeq -P^\theta(n,0,n).
\end{equation}
The formulas and relationships for $\theta(0,n,n)$ are very similar.
The above numerical findings indicate that the most dominant entropy transfers to the $\theta(0,0,2n)$ mode occur from the $\theta(n,0,n)$ and $\theta(0,n,n)$. Also, the approximate relationship $S(n,0,n|0,0,2n|-n,0,n)  \simeq -P^\theta(n,0,n)$ and the equivalent relationship for the $\theta(0,n,n)$ mode yield
\begin{equation}
\theta(0,0,2n) \simeq -\frac{1}{2n\pi}
\end{equation}
that matches quite well with the simulation data for $P=6.8$ and 0.2 (listed in Table~\ref{tab:theta002n}).  Using the above result we can immediately derive 
\begin{equation}
E^\theta(0,0,2n) \simeq \frac{1}{4\pi^2 n^2}
\end{equation}
that matches well with the upper branch of the entropy spectrum as shown in Figs.~\ref{thermspec_pr0p02},~\ref{thermspec_pr0p2},~\ref{thermspec_pr1}, and ~\ref{thermspec_pr6p8}.  The lower branch of the entropy spectrum corresponds to modes other than $\theta(0,0,2n)$.   The dual branches appear to arise due to the free-slip boundary conditions, and they have been observed in the simulation by Vincent and Yuen~\cite{Vincent:PRE_1999} and Paul {\em et al.}~\cite{Paul:2Dspectrum}.  Note that the dual branches in the entropy spectrum have not been reported for no-slip~\cite{Kerr:1996} and periodic boundary conditions~\cite{borue:1997,skandera:2008}.

The above arguments that support $E^\theta(0,0,2n) \sim 1/n^2$ is essentially numerical and phenomenological that works for $P=6.8$ and 1.  For lower Prandtl numbers, $T^\theta(n,0,n)$ is not approximately equal to $S(n,0,n|0,0,2n|-n,0,n)$ possibly due to significant entropy transfers to other modes, or due to thermal diffusion.  Note however that the above quantities are within a factor of two, consequently $\theta(0,0,2n) \simeq -\frac{1}{2n\pi}$ holds even for lower Prandtl number within a factor of two.  

The dual branches in the entropy spectrum adds complications to the energy fluxes discussed in the paper.  The temperature modes on the upper branch have significantly higher entropy, but they are only a few in numbers.  Hence, the nonlinear energy transfers arising from the upper branch is possibly insignificant.  The number of modes involved in the lower branch is quite large, and they are likely to provide the energy flux.

{\bf Acknowledgements:} 
We thank Krishna Kumar, Supriyo Paul, and Stephan Fauve for valuable discussions and suggestions.   We thank Computational Research Laboratories (CRL), Pune for providing access to the supercomputer EKA where the above simulations were performed.  This work was supported by funds from Department of Science and Technology, India as Swarnajayanti fellowship to MKV.  
\pagebreak

\pagebreak
\begin{center}
{\bf FIGURES}
\end{center}


   \newpage
  
   \newpage
  
   \newpage
 
  \newpage 
  
   \newpage
  
   \newpage

 \newpage
  
   \newpage
  
   \newpage

 \newpage
  
   \newpage
  
   \newpage
  
   \newpage
  
   \newpage

 \newpage
\begin{center}
{\bf TABLES}
\end{center} 
 
  \begin{table}[ht]
\caption{Estimates of the viscous dissipation ($\epsilon^u$) and the thermal diffusion rates ($\epsilon^\theta$)  from numerical simulation and by using theoretical relationships [Eqs.~(\ref{eq:Piu_exact}-\ref{eq:Pitheta_exact})], inverse of the Bolgiano length ($l_B^{-1}$), Kolmogorov's dissipative wavenumber ($k_d$),  and Kolmogorov's diffusive wavenumber ($k_c$).  The reported quantities are nondimensional: $\epsilon^u = (Nu-1)/\sqrt{RP}$, $\epsilon^\theta = Nu/\sqrt{RP}$, $k_d = [(Nu-1)R/P^2]^{1/4}$, and $k_c = [(Nu-1)RP]^{1/4}$. }
\vspace{1cm}
\begin{tabular}{ c c c c c c c c c c c} \hline
$P$ & $ R $&  $Nu$ & $\epsilon^{u}$ & $\epsilon^{u}$ & $\epsilon^{\theta}$ & $\epsilon^{\theta}$ &$l_{B}^{-1}$ & $k_d$ & $k_c$ \\
    &      &        & (estim).      &   (comp.)       &    (estim.)        &   (comp.)            &             &       &  \\ \hline
0.02& $2.6\times10^{6}$& 8.5 & 0.033 & 0.032 & 0.037 &0.037 & 5.2 &470.3 &25.0\\ 
0.2 & $6.6\times10^{6}$& 17 & 0.014 & 0.014 & 0.015 & 0.015& 8.2 &227.0 &68.0\\ 
1.0 & $6.6\times10^{6}$ & 32 & 0.012 & 0.0082 & 0.013 &0.0085 &9.0 & 108.6 &108.6\\ 
6.8 & $6.6\times10^{6}$ & 30 & 0.004 & 0.0043 & 0.0042 &0.0042 &15.0& 44.2 &186.2\\ \hline
\end{tabular}
\label{Table1}
\end{table}

\begin{table}[ht]
\caption{Numerical values of $\theta(0,0,2)$, $\theta(0,0,4)$, $\theta(0,0,6)$, and $\theta(0,0,8)$ modes for $P=6.8,0.2$ and 0.02.  Our phenomenological arguments with numerical ingredients indicate that $\theta(0,0,2n) \simeq -1/(2n\pi)$.  }
\vspace{1cm}
\begin{tabular}{ c c c c c } \\ \hline
$P$& $\theta(0,0,2)$& $\theta(0,0,4)$ & $\theta(0,0,6)$&$\theta(0,0,8)$ \\ \hline
6.8& -0.16  &   -0.077 &  -0.050 & -0.036 \\ 
0.2 & -0.15 &  -0.061  &  -0.031  & -0.017\\
0.02& -0.13 & -0.040 &   -0.017  & -0.0081 \\ \hline
-$\frac{1}{2n\pi}$& -0.16 &  -0.080 &  -0.053 & -0.040 \\ \hline
\end{tabular}
\label{tab:theta002n}
\end{table}

  \begin{table}[ht]
\caption{For high-P ($P=6.8$) and low-P ($P=0.2$), the numerical values of the nonlinear entropy transfer rates $T^\theta$,  entropy production rates $P^\theta$, and the nonlinear entropy transfer rates $S({\bf k|p|q})$ from the mode $\theta(0,0,2n)$ mode to the modes $\theta(n,0,n)$ or $\theta(0,n,n)$.}
\vspace{1cm}
\begin{tabular}{ c| c c c|c c c} \hline
    &            &   $ P=6.8$  &                      &                &     $P=0.2$         &                           \\ \hline
mode&$T^{\theta}$& $S({\bf k|p|q})$ &$P^\theta$& $T^{\theta}$&  $S({\bf k|p|q})$ &$P^\theta$ \\ \hline
(1,0,1)& -1.2e-4 &-1.1e-4  &1.1e-4  & -1.3e-3 & -2.7e-3& 2.9e-3\\ 
(0,1,1)& -1.0e-7 & -7.5e-8 & 7.5e-8& -3.1e-4 & -3.4e-4& 3.6e-4\\
(2,0,2)&-7.0e-7& -5.0e-7&5.5e-7 & -2.5e-6 & -2.6e-5 & 3.4e-5 \\
(0,2,2)&-1.6e-6 & -1.2e-6 & 1.3e-6  &-6.1e-5 & -4.9e-5 &6.3e-5 \\
(3,0,3)&3.0e-7 &2.1e-7 &-2.2e-7 &-7.0e-7 & -1.4e-6 & 2.3e-6  \\
(0,3,3)&-1.0e-7 & -9.3e-8  &1.0e-7 &-5.7e-6 & -5.3e-6 & 9.2e-6 \\
(4,0,4)&3.0e-7 & -3.2e-7 & 3.5e-7 &-8.0e-7 & -1.0e-6 &2.4e-6 \\
(0,4,4)& -4.0e-7 &-4.0e-7 &3.9e-7  & 1.0e-6 & -7.0e-7  & -1.6e-6 \\  \hline
\end{tabular}
\label{tab:Ttheta}
\end{table}
  
\end{document}